\documentclass[twoside,11pt]{article}

\usepackage{verbatim}
\usepackage{caption}
\usepackage{subcaption}
\usepackage{fancyvrb}
\usepackage[T1]{fontenc}
\usepackage{hyperref}
\usepackage{array}

\usepackage{jmlr2e}
\usepackage{amsmath}

\jmlrheading{1}{2016}{1-29}{12/14; Revised 3/16}{}{Julie Josse and Stefan Wager}
\ShortHeadings{Regularized Low-Rank Matrix Estimation}{Josse and Wager}
\firstpageno{1}

\editor{Rob Fergus}

\newcommand{\shrink}{{\operatorname{shrink}}}

\usepackage{verbatim}
\usepackage{caption}
\usepackage{subcaption}
\usepackage{color}
\usepackage{fancyvrb}
\usepackage{algorithm}
\usepackage{algorithmic}

\usepackage{hyperref}

\newcommand{\iter}{{\operatorname{iter}}}
\newcommand{\stable}{{\operatorname{stable}}}

\usepackage{./StefanTexCommandsV2}

\graphicspath{{./}}

\newtheorem{prop}{Proposition}

\newtheorem{theo}[prop]{Theorem}

\graphicspath{{./pic/}}

\title{Bootstrap-Based Regularization for \\ Low-Rank Matrix Estimation}

\author{Julie Josse \email josse@agrocampus-ouest.fr \\
\addr Department of Applied Mathematics, Agrocampus Ouest, Rennes, France, \\
INRIA Saclay, Universit\'e Paris-Sud, Orsay, France.
\AND
\name Stefan Wager \email swager@stanford.edu \\
\addr Department of Statistics, Stanford University, Stanford, U.S.A.}

\newcommand{\SPACEBIG}{1}
\newcommand{\SPACESMALL}{1}

\def\spacingset#1{\renewcommand{\baselinestretch}%
{#1}\small\normalsize} \spacingset{1}

\begin{document}

\clearpage\thispagestyle{empty}
\maketitle

\begin{abstract}
We develop a flexible framework for low-rank matrix estimation that allows us to transform noise models into regularization schemes via a simple bootstrap algorithm. Effectively, our procedure seeks an autoencoding basis for the observed matrix that is stable with respect to the specified noise model; we call the resulting procedure a \emph{stable autoencoder}. In the simplest case, with an isotropic noise model, our method is equivalent to a classical singular value shrinkage estimator. For non-isotropic noise models---e.g., Poisson noise---the method does not reduce to singular value shrinkage, and instead yields new estimators that perform well in experiments. Moreover, by iterating our stable autoencoding scheme, we can automatically generate low-rank estimates without specifying the target rank as a tuning parameter.
\end{abstract}

\spacingset{\SPACEBIG}

\section{Introduction}

Low-rank matrix estimation plays a key role in many scientific and engineering tasks, including collaborative filtering \citep{koren2009matrix}, genome-wide studies \citep{leek2007capturing,price2006principal}, and magnetic resonance imaging \citep{candes2013sure,lustig2008compressed}.
Low-rank procedures are often motivated by the following statistical model. Suppose that we observe a noisy matrix $X \in \RR^{n \times p}$ drawn from some distribution $\law(\mu)$ with $\EE[\mu]{X} = \mu$, and that we have scientific reason to believe that $\mu$ admits a parsimonious, low-rank representation. Then, we can frame our statistical goal as trying to recover the underlying $\mu$ from the observed $X$. \Citet{d2012approximation},
\citet{candes2010power}, \citet{Chat2014univ}, \citet{gavish2014optimalshrink}, \citet{shabalin2013reconstruction}, and others have studied regimes where it is possible to accurately do so.

\paragraph{Singular-value shrinkage}
Classical approaches to estimating $\mu$ from $X$ are centered around singular-value decomposition (SVD) algorithms. Let
\begin{equation}
\label{eq:svd}
X = \sum_{l = 1}^{\min\{n, \, p\}} u_l \, d_l \, v_l^\top
\end{equation}
denote the SVD of $X$.%\footnote{Recall that, in the SVD, the $\{u_l\}$ and $\{v_l\}$ form orthogonal sets, and the $d_l$ form a decreasing non-negative sequence.}
Then, if we believe that $\mu$ should have rank $k$, the standard SVD estimator $\hmu_k$ for $\mu$ is
\begin{equation}
\label{eq:svd_mu}
\hmu_k = \sum_{l = 1}^{k} u_l \, d_l \,  v_l^\top.
\end{equation}
In other words, we estimate $\mu$ using the closest rank-$k$ approximation to $X$.
Often, however, the plain rank-$k$ estimator \eqref{eq:svd_mu} is found to be noisy, and its performance can be improved by regularization. Existing approaches to regularizing $\hmu_k$ focus on singular value shrinkage, and use
\begin{equation}
\label{eq:svd_shrink}
%\hmu_k^\shrink = \sum_{l = 1}^{k} u_l \, \psi\p{d_l} \, v_l^\top, 
\hmu^\shrink =\sum_{l = 1}^{\min\{n, \, p\}} u_l \, \psi\p{d_l} \, v_l^\top, 
\end{equation}
where $\psi$ is a shrinkage function that is usually chosen in a way that makes $\hmu^\shrink$ the closest to $\mu$ according to a loss function. 
Several authors have proposed various choices for $\psi$ \citep[e.g.,][]{candes2013sure,Chat2014univ,gavish2014optimal,josse2014adaptive,shabalin2013reconstruction,verbanck2013regularised}.

Methods based on singular-value shrinkage have achieved considerable empirical success. They also have provable optimality properties in the Gaussian noise model where $X = \mu + \varepsilon$ and the $\varepsilon_{ij}$ are independent and identically distributed Gaussian noise terms \citep{shabalin2013reconstruction}. However, in the non-Gaussian case, mere singular-value shrinkage can prove to be limiting, and we may also need to rotate the singular vectors $u_l$ and $v_l$ in order to achieve good performance.

\paragraph{Stable autoencoding}
In this paper, we propose a new framework for regularized low-rank estimation that does not start from the singular-value shrinkage point of view. Rather, our approach is motivated by a simple plug-in bootstrap idea \citep{efron1993introduction}. It is well known that the classical SVD estimator $\hmu_k$ can be written as \citep{bourlard1988auto,baldi1989neural}
\begin{equation}
\label{eq:autoencoder}
\hmu_k = XB_k, \where B_k = \argmin_B \left\{\Norm{X - XB}_2^2 : \rank{B} \leq k\right\},
\end{equation}
where $\Norm{M}_2^2 = \tr\p{M^\top M}$ denotes the Frobenius norm.
The matrix $B$, called a linear \emph{autoencoder} of $X$, allows us to encode the features of $X$ using a low-rank representation. 

Now, in the context of our noise model $X \sim \law(\mu)$, we do not just want to compress $X$, and instead want to recover $\mu$ from $X$. From this perspective, we would much prefer to estimate $\mu$ using an oracle encoder matrix that formally provides the best linear approximation of $\mu$ given our noise model
\begin{equation}
\label{eq:autoencoder_oracle}
\hmu_k^* = XB_k^*, \where B_k^* = \argmin_B \left\{\EE[X \sim \law(\mu)]{\Norm{\mu - XB}_2^2} : \rank{B} \leq k\right\}.
\end{equation}
We of course cannot solve for $B_k^*$ because we do not know $\mu$.
But we can seek to approximate $B_k^*$ by solving the optimization problem
\eqref{eq:autoencoder_oracle} on a well-chosen bootstrap distribution.

More specifically, our goal is to create bootstrap samples $\tX \sim \tlaw(X)$
such that the distribution of $\tX$ around $X$ is representative of the
distribution of $X$ around $\mu$. Then, we can solve an analogue to
\eqref{eq:autoencoder_oracle} on the bootstrap samples $\tX$:
\begin{equation}
\label{eq:autoencoder_boot}
\hmu_k^\stable = X\hB_k, \where \hB_k = \argmin_B \left\{\EE[\tX \sim \tlaw(X)]{\Norm{X - \tX B}_2^2} : \rank{B} \leq k\right\}.
\end{equation}
We call this choice of $\hB_k$ a \emph{stable autoencoder} of $X$,
as it provides a parsimonious encoding of the features of $X$ that
is stable when perturbed with bootstrap noise $\tlaw(X)$.
The motivation behind this approach is that we want to shrink
$\hmu_k$ aggressively along directions where the boostrap reveals
instability, but do not want to shrink $\hmu_k$ too much along
directions where our measurements are already accurate.

\paragraph{Bootstrap models for stable autoencoding}
A challenge in carrying out the program \eqref{eq:autoencoder_boot} is in
choosing how to generate the bootstrap samples $\tX$. In the classical statistical
setting, we have access to $m \gg 1$ independent training samples and can thus
create a bootstrap dataset by simply re-sampling the training data with replacement.
In our setting, however, we only have a single matrix $X$, i.e., $m = 1$. Thus,
we must find another avenue for creating bootstrap samples $\tX$.

To get around this limitation, we use a \emph{L\'evy bootstrap}
\citep{wager2016data}. Before defining the abstract bootstrap scheme below,
we first discuss some simple examples.
In the {\bf Gaussian} case $X_{ij} \sim \nn\p{\mu_{ij}, \, \sigma^2}$, L\'evy
boostrapping is equivalent to a parametric bootstrap:
\begin{equation}
\label{eq:gauss_noise}
\tX_{ij} = X_{ij} +\teps_{ij}, \ \where \ \teps_{ij} \simiid \nn\p{0, \, \frac{\delta}{1 - \delta} \,  \sigma^2},
\end{equation}
and $\delta \in \p{0, \, 1}$ is a tuning parameter
that governs the regularization strength.
Meanwhile, in the {\bf Poisson} case $X_{ij} \sim \text{Poisson}\p{\mu_{ij}}$,
L\'evy bootstrapping involves randomly deleting a fraction $\delta$ of
the counts comprising the matrix $X$, and up-weighting the rest:
\begin{equation}
\label{eq:poisson_noise}
\tX_{ij} \sim \frac{1}{1 - \delta} \, \text{Binomial}\p{X_{ij}; \, 1 - \delta},
\end{equation}
where again $\delta \in (0, \, 1)$ governs the regularization strength.
In the case $\delta = 0.5$, this is equivalent to the ``double-or-nothing''
bootstrap on the individual counts of $X$ \citep[e.g.,][]{owen2012bootstrapping}.

These two examples already reveal a variety of different phenomena.
On one hand, stable autoencoding with the Gaussian noise model
\eqref{eq:gauss_noise} reduces to a singular-value shrinkage estimator
(Section \ref{sec:boot}), and thus leads us back to the classical literature
on low-rank matrix estimation.
Conversely, with the Poisson-adapted bootstrap \eqref{eq:poisson_noise},
the method \eqref{eq:autoencoder_boot} rotates singular vectors instead of
just shrinking singular vales; and in our experiments, it
outperforms several variants of singular-value shrinkage that have
been proposed in the literature.

\paragraph{The L\'evy bootstrap}
Having first surveyed our two main examples above, we now present the more
general L\'evy bootstrap that can be used to carry out stable autoencoding
in a wider variety of exponential family models.
To motivate this approach, suppose that we can generate
our matrix $X$ as a sum of independent components:
\begin{equation}
\label{eq:levy_decomp}
X = \sum_{b = 1}^B X^{(B)}_b, \ \where \ X^{(B)}_1, \, ..., \, X^{(B)}_B
\end{equation}
are independent and identically distributed. For example, if $X$ is
Gaussian with $X_{ij} \sim \nn\p{\mu_{ij}, \, \sigma}$, then \eqref{eq:levy_decomp}
holds with \smash{$(X^{(B)}_b)_{ij} \sim \nn\p{\mu_{ij}/B, \, \sigma/B}$}.
If the above construction were to hold and we also knew the individual
components \smash{$X_b^{(B)}$}, we could easily create bootstrap 
samples $\tX$ as
\begin{equation}
\label{eq:subs_boot}
\tX = \frac{1}{1 - \delta} \, \sum_{b = 1}^B \tW_b \, X^{(B)}_b, \ \where \ \tW_b \,\simiid\, \text{Bernoulli}\p{1 - \delta},
\end{equation}
and $\delta \in (0, \, 1)$ governs the noising strength. Now in reality,
we do not know the terms in \eqref{eq:levy_decomp}, and so cannot carry out
\eqref{eq:subs_boot}.

However, \citet{wager2016data} establish conditions under which this
limitation does not matter.
Suppose that $X$ is drawn from an exponential family distribution
\begin{equation}
\label{eq:expfam}
X \sim f_\theta(\cdot), \ \ f_\theta(x) = h(X) \, \exp\left[\sum_{i, \, j} \Theta_{ij} X_{ij} - \zeta\p{\Theta}\right],
\end{equation}
where $\Theta \in \RR^{n \times p}$ is an unknown parameter vector,
$h(\cdot)$ is a carrier distribution, and $\zeta(\cdot)$ is the log-partition function.
Suppose, moreover, that $X$ has an infinitely divisible distribution, or,
equivalently, that $X = A(1)$ for some matrix-valued L\'evy process
$A(t) \in \RR^{n \times p}$ for $t > 0$ \citep[e.g.,][]{durrett2010probability}.
Then, we can always generate bootstrap replicates $\tX$ distributed as
\begin{equation}
\label{eq:levy_boot}
\tX \sim \tlaw_\delta(X) := \frac{1}{1 - \delta} \, A(1 - \delta) \cond A(1) = X, \ \text{ for any } \delta \in (0, \, 1),
\end{equation}
without requiring knowledge of the underlying $\Theta$; \citet{wager2016data}
provide explicit formulas for carrying out \eqref{eq:levy_boot} that
only depend on the carrier $h(\cdot)$.

The upshot of this bootstrap scheme is that it allows us to preserve the
generative structure encoded by $\Theta$ without needing to know the true parameter.
In the Gaussian and Poisson cases, \eqref{eq:levy_boot} reduces to
\eqref{eq:gauss_noise} and \eqref{eq:poisson_noise} respectively;
however, this L\'evy bootstrap framework also induces other
noising schemes, such as multiplicative noising when $X$ has a Gamma distribution.

\paragraph{Rank selection via iterated stable autoencoding}

Returning now to our main focus, namely regularized low-rank
matrix estimation, we note that one difficulty with the
estimator $\hmu_k^\stable$ \eqref{eq:autoencoder_boot}
is that we need to select the rank $k$ of
the estimator beforehand in addition to the shrinkage parameter
$\delta$. Surprisingly, we can get around this issue by
iterating the optimization problem \eqref{eq:autoencoder_boot} until
we converge to a limit:
\begin{equation}
\label{eq:autoencoder_freerank}
\hmu^\iter = X\hB, \where \hB = \argmin_B  \left\{\EE[\tX  \sim \tlaw\p{\hmu^\iter, \, X}]{\Norm{\hmu^\iter  -  \tX  B}_2^2} \right\}.
\end{equation}
In Section \ref{sec:iter}, we establish conditions under which the
iterative algorithm implied above in fact converges and, moreover,
the resulting fixed point $\hmu^\iter$ is low rank.
In our experiments, this \emph{iterated stable autoencoder} does a
good job at estimating $k$ of the underlying signal; thus, all the
statistician needs to do is to specify a single regularization parameter
$\delta \in (0, \, 1)$ that simultaneously controls both the amount of
shrinkage and the rank $k$ of the final estimate $\hmu$.

To summarize, our approach as instantiated in
\eqref{eq:autoencoder_boot} and \eqref{eq:autoencoder_freerank} provides
us with a flexible framework for transforming noise models $\law(\cdot)$
into regularized matrix estimators via the L\'evy bootstrap.
In the Gaussian case, our framework
yields estimators that resemble best-practice singular-value shrinkage
methods. Meanwhile, in the non-Gaussian case, stable autoencoding allows
us to learn new singular vectors for $\hmu$. In our experiments, this
allowed us to substantially improve over existing techniques.

Finally, we also discuss extensions to stable autoencoding: In Section \ref{sec:ca},
we show how to use our method to regularize correspondence analysis
\citep{green1984ca,green2007ca}, which is one of the most popular ways
to analyze multivariate count data and underlies several modern machine
learning algorithms.

A software implementation of the proposed methods is available through the
\texttt{R}-package \texttt{denoiseR} \citep{josse2016denoiser}.

\subsection{Related work}

There is a well-known duality between regularization and feature noising schemes. As shown by \citet{bishop1995training}, linear regression with features perturbed with Gaussian noise, i.e.,
$$\hbeta = \argmin_\beta\left\{\EE[\varepsilon_{ij} \simiid \nn\p{0, \, \sigma^2}]{\Norm{Y - \p{X + \varepsilon}\beta}_2^2}\right\}, $$
is equivalent to ridge regularization with Lagrange parameter $\lambda = n\sigma^2$: 
$$ \hbeta^{(R)}_\lambda = \argmin_\beta \left\{\Norm{Y - X\beta} + \lambda \Norm{\beta}_2^2 \right\}. $$
Because ridge regression is equivalent to adding homoskedastic noise to $X$, we can think of ridge regression as making the estimator robust against round perturbations to the data. 

However, if we perturb the features $X$ using non-Gaussian noise or
are working with a non-quadratic loss function, artificial feature noising
can yield new regularizing schemes with desirable properties
\citep{globerson2006nightmare,simard2000transformation,van2013learning,wager2013dropout,wang2013feature}. 
Our proposed estimator $\hmu_k^\stable$ can be seen as an addition to this
literature, as we seek to regularize $\hmu_k$ by perturbing the autoencoder
optimization problem. The idea of regularizing via feature noising is also
closely connected to the dropout learning algorithm for training neural networks
\citep{srivastava2014dropout}, which aims to regularize a neural network by
randomly omitting hidden nodes during training time. Dropout and its
generalizations have been found to work well in many large-scale prediction tasks
\citep[e.g.,][]{baldi2014dropout,goodfellow2013maxout,krizhevsky2012imagenet}.

Our method can be interpreted as an empirical Bayes
estimator \citep{efron2012large,robbins1985empirical}, in that
the stable autoencoder problem \eqref{eq:autoencoder_boot} seeks to find the
best linear shrinker in an empirically chosen Bayesian model. There is also
an interesting connection between stable autoencoding, and more traditional
Bayesian modeling such as latent Dirichlet allocation (LDA) \citep{blei2003latent},
in that the L\'evy bootstrap \eqref{eq:levy_boot} uses a generalization of the LDA
generative model to draw bootstrap samples $\tX$. Thus, our method can be seen as
benefiting from the LDA generative structure without committing to full Bayesian
inference \citep{kucukelbir2015population,wager2014altitude,wager2016data}.

There is a large literature on low-rank exponential family estimation
\citep{collins2001ageneralization,deLeeuw2006PCA,fithian2013scalable,good1985analysis,li2013simple}.
In the simplest form of this idea, each matrix entry is modeled using
the generic exponential family distribution \eqref{eq:expfam};
the goal is then to maximize the log-likelihood of $X$
subject to a low-rank constraint on the natural parameter matrix $\Theta$,
rather than the mean parameter matrix $\mu$ as in our setting.
The main difficulty with this approach is that the resulting rank-constrained
problem is no longer efficiently solvable; one way to avoid this issue is to relax
the rank constraint into a nuclear norm penalty.
Extending our bootstrap-based regularization framework to low-rank
exponential family estimation would present an interesting avenue for further work.
We also note the work of \citet{buntine2002variational}, who seeks to maximize
the multinomial log-likelihood of $X$ subject to a low-rank constraint on $\mu$
using an approximate variational method.

Finally, one of the advantages of our stable autoencoding approach is that it lets us
move beyond singular value shrinkage, and learn better singular vectors than those
provided by the SVD. Another approach to way to improve on the quality of the learned
singular vectors is to impose structural constraints on them, such as sparsity
\citep{jolliffe2003modified,udell2014generalized,witten2009penalized,zou2006sparsepca}.

\section{Fitting stable autoencoders}
\label{sec:boot}

In this section, we show how to solve \eqref{eq:autoencoder_boot} under various bootstrap models $\tlaw(\cdot)$. This provides us with estimators $\hmu^\stable$ that are interesting in their own right, and also serves as a stepping stone to the iterative solutions from Section \ref{sec:iter} that do not require pre-specifying the rank $k$ of the underlying signal. 

\paragraph{Isotropic stable autoencoders and singular-value shrinkage}
% \label{sec:gaussianautoenco}

At first glance, the estimator $\hmu_k^\stable$ defined in \eqref{eq:autoencoder_boot} may seem like a surprising idea. It turns out, however, that under the isotropic Gaussian\footnote{Theorem \ref{theo:gauss} holds for all isotropic noise models with $\Var{\varepsilon_{ij}} = \sigma^2$ for all $i$ and $j$, and not just the Gaussian one. However, in practice, isotropic noise is almost always modeled as Gaussian.} noise model
\begin{equation}
\label{eq:gauss}
X = \mu + \varepsilon, \with \varepsilon_{ij} \simiid \nn\p{0, \, \sigma^2} \text{ for all } i = 1, \, ..., \, n \eqand j = 1, \, ..., \, p,
\end{equation}
$\hmu_k^\stable$ with bootstrap noise as in \eqref{eq:gauss_noise} is equivalent to a classical singular-value shrinkage estimator \eqref{eq:svd_shrink} with $\psi(d) = d/(1 + \lambda/d^2)$ and $\lambda = \delta/(1 - \delta) \, n \sigma^2$.

\begin{theo}
\label{theo:gauss}
Let $\hmu_k^\stable$ be the rank-$k$ estimator for $\mu$ induced by the stable autoencoder \eqref{eq:autoencoder_boot} with a bootstrap model $\tlaw_\delta(\cdot)$ as defined in \eqref{eq:gauss_noise} with some $0 < \delta < 1$. This estimator can also be written as the solution to a ridge-regularized autoencoder:
\begin{equation}
\label{eq:ridge}
\hmu^{\stable}_k =  X \hB_k, \; \where \; \hB_k = \argmin_B \left\{\Norm{X - XB}_2^2 + \lambda \Norm{B}_2^2 : \rank{B} \leq k\right\}
\end{equation}
with $\lambda = \delta/(1 - \delta) \, n \sigma^2$. Moreover, using notation from \eqref{eq:svd}, we can write $\hmu_k^\stable$ as
\begin{equation}
\label{eq:gauss_thm}
\hmu_k^\stable = \sum_{l = 1}^k u_l \, \frac{d_l}{1 + \lambda/d_l^2} \, v_l^\top.
\end{equation}
\end{theo}

In the isotropic Gaussian noise case, singular-value shrinkage methods were shown by \citet{shabalin2013reconstruction} to have strong optimality properties for estimating $\mu$. Thus, it is reassuring that our framework recovers an estimator of this class in the Gaussian case. In fact, the induced shrinkage function resembles a first-order approximation to the one proposed by \citet{verbanck2013regularised}.

\paragraph{Non-isotropic stable autoencoders}
%\label{sec:poiss}

Stable autoencoders with isotropic noise are attractive in the sense that we can carefully analyze their behavior in closed form.
However, from a practical point of view, our procedure is most useful outside of the isotropic regime, as it induces new estimators $\hmu$ that do not reduce to singular-value shrinkage. Even in the non-isotropic noise model, low-rank stable autoencoders can still be efficiently solved, as shown below.

\begin{theo}
\label{theo:compute}
For a generic bootstrap model $\tlaw(\cdot)$, the matrix $\hB_k$ from \eqref{eq:autoencoder_boot} can be obtained as follows:
\begin{equation}
\label{eq:nongauss}
\hB_k = \argmin_B \left\{\Norm{X - XB}_2^2 + \Norm{S^{\frac12}B}_2^2 : \rank{B} \leq k\right\},
\end{equation}
where $S$ is a $p \times p$ diagonal matrix with
\begin{equation}
\label{eq:S}
S_{jj} = \sum_{i = 1}^n \Var[\tX \sim \tlaw\p{X}]{\tX_{ij}}.
\end{equation}
From a computational point of view, we can write the solution $\hB_k$ of \eqref{eq:nongauss} as
\begin{align}
\label{eq:nongauss_svd}
&\hB_k = \argmin_B \left\{ \tr\p{\p{B - \hB}^\top \p{X^\top X + S} \p{B - \hB}} : \rank{B} \leq k\right\}, \where \\
\label{eq:closed_form}
&\hB = \p{X^\top X + S}^{-1} X^\top X
\end{align}
is the solution of \eqref{eq:nongauss} without the rank constraint.
\end{theo}

The optimization problem in \eqref{eq:nongauss_svd} can be easily solved by taking the top $k$ terms from the eigenvalue decomposition of $\hB^\top \p{X^\top X + S} \hB$; the matrix $\hB_k$ can then be recovered by solving a linear system \citep[e.g.,][]{takane2013gsvd}. Thus, despite what we might have expected, solving the low-rank constrained stable autoencoder problem  \eqref{eq:autoencoder_boot} with a generic noise model is not substantially more computationally demanding than singular-value shrinkage. 
Note that in \eqref{eq:closed_form}, the matrix $S$ is not equal to a constant times the identity matrix due to the non-isotropic noise, and so the resulting singular vectors of $\hmu_k^\stable=X\hB_k$ are not in general the same as those of $X$. 

\paragraph{Selecting the tuning parameter}

Our stable autoencoder depends on a tuning parameter $\delta \in (0, \, 1)$, corresponding to
the fraction of the information in the full data $X$ that we throw away when creating pseudo-datasets
$\tX$ using the L\'evy bootstrap \eqref{eq:levy_boot}. More prosaically, the parameter $\delta$ manifests itself
as a multiplier $\delta / (1 - \delta) \in (0, \, \infty)$ on the effective stable autoencoding penalty,
either explicitly in \eqref{eq:ridge} or implicitly in \eqref{eq:S} through the dependence on $\tlaw_\delta(\cdot)$.

One plausible default value is to set $\delta = 1/2$. This corresponds to using half of the information in the full
data $X$ to generate each bootstrap sample $\tX$, and is closely related to bagging \citep{breiman1996bagging};
see \citet{buja2006observations} for a discussion of the connections between half-sampling and bagging.
Conversely, we could also opt for a data-driven choice of $\delta$. The software implementation of stable
autoencoding in \texttt{denoiseR} \citep{josse2016denoiser} provides a cell-wise cross-validation algorithm
for picking $\delta$.

Finally, we note that our estimator is not in general invariant to transposition $X \rightarrow X^\top$. For example,
in the isotropic case \eqref{eq:ridge}, we see that $\lambda$ depends on $n$ but not on $p$. Meanwhile, in the
non-isotropic case \eqref{eq:nongauss}, transposition may also affect the learned singular vectors. By default, we
transpose $X$ such that $n > p$, i.e., we pick the transposition of $X$ that makes the matrix $B$ smaller.

\section{Iterated stable autoencoding}
\label{sec:iter}

One shortcoming of the stable autoencoders discussed in the previous section is that we need to specify the rank $k$ as a tuning parameter. Selecting the rank for multivariate methods is often a difficult problem, and many heuristics are available in the literature \citep{jolliffe2002pca, josse2011gcvpca}. The stable autoencoding framework, however, induces a simple solution to the rank-selection problem: As we show here, iterating our estimation scheme from the previous section automatically yields low-rank solutions, and allows us to specify a single tuning parameter $\delta$ instead of both $\delta$ and $k$.

\spacingset{\SPACESMALL}
\begin{algorithm}[t]
\caption{Low-rank matrix estimation via iterated stable autoencoding.}
\label{alg:iter}
\begin{algorithmic}
\STATE $\hmu \gets X$
\STATE $S_{jj} \gets \sum_{i = 1}^n \Var[\tX \sim \tlaw_\delta\p{X}]{\tX_{ij}}$ for all $j = 1, \, ..., \, p$
\WHILE{algorithm has not converged}
  \STATE $\hB \gets \p{\hmu^\top \, \hmu + S}^{-1} \hmu^\top \, \hmu$
  \STATE $\hmu \gets X \, \hB$
\ENDWHILE
\end{algorithmic}
\end{algorithm}
\spacingset{\SPACEBIG}

At a high level, our goal is to find a solution to
\begin{equation}
\label{eq:iter_target}
\hmu^\iter = X\hB, \where \hB = \argmin_B  \left\{\EE[\tX  \sim \tlaw_\delta\p{\hmu^\iter, \, X}]{\Norm{\hmu^\iter  -  \tX  B}_2^2} \right\} 
\end{equation}
by iteratively updating $\hB$ and $\hmu$. As seen in the previous section, stable autoencoding only depends
on $\tlaw_\delta\p{\hmu, \, X}$ through the first two moments of $\tX$; here, we simply specify them as
\begin{equation}
\EE[\tX  \sim \tlaw_\delta\p{\hmu, \, X}]{\tX} = \hmu \ \eqand \ \Var[\tX  \sim \tlaw_\delta\p{\hmu, \, X}]{\tX} = \Var[\tX  \sim \tlaw_\delta\p{X}]{\tX},
\end{equation}
where $\tlaw_\delta\p{X}$ is obtained using the L\'evy bootstrap as before.

Now, using the unconstrained solution \eqref{eq:closed_form} from Theorem \ref{theo:compute}
to iterate on the relation \eqref{eq:iter_target}, we get the formal procedure described in Algorithm \ref{alg:iter}.
Note that we do not update the matrix $S$, which encodes the variance of the noise distribution,
and only update $\hmu$. As shown below, our algorithm converges to a well-defined solution;
moreover, the solution is regularized in that $\hmu^\top \hmu$ is smaller than $X^\top X$
with respect to the positive semi-definite cone ordering.

\begin{theo}
\label{theo:iterate}
Algorithm \ref{alg:iter} converges to a fixed point $\hmu = X\hB$. Moreover,
$$ \hmu^\top\hmu \preceq X^\top X. $$
\end{theo}

Moreover, iterated stable autoencoding can provide generic low-rank solutions $\hmu$.

\begin{theo}
\label{theo:low_rank}
Let $\hmu$ be the limit of our iterative algorithm, and let $u \in \RR^p$ be any (normalized) eigenvector of $\hmu^\top\hmu$. Then, either
$$ \Norm{\hmu \, u}_2 = 0, \; \text{or} \; \Norm{\hmu \, u}_2 \geq \frac{1}{\Norm{X S^{-1} u}_2^2}. $$
\end{theo}

The reason our algorithm converges to low-rank solutions is that our iterative scheme does not have any fixed points ``near'' low-dimensional subspaces. Specifically, as shown in Theorem \ref{theo:low_rank}, for any eigenvector of $\hmu^\top\hmu$, either $\Norm{\hmu \, u}_2$ must be larger than some cutoff, or it must be exactly zero. Thus, $\hmu$ cannot have any small but non-zero singular values. In our experiments, we have found that our algorithm in fact conservatively estimates the true rank of the underlying signal.

Finally we note that, in the isotropic case, our iterative algorithm again admits a closed-form solution. Looking at this solution can give us more intuition about what our algorithm does in the general case. In particular, we note that the algorithm never shrinks a singular value by more than a factor $1/2$ without pushing it all the way to 0.

\begin{prop}
\label{prop:iter_gauss}
In the isotropic Gaussian case \eqref{eq:gauss} with $\delta = 1/2$, our iterative algorithms converges to
\begin{equation}
\label{eq:iter_gauss}
\hmu^{\iter} = \sum_{l = 1}^{\min\{n, \, p\}} u_l \, \psi\p{d_l} \, v_l^\top, \; \where \; \psi\p{d} =
\begin{cases}
\frac12 \p{d + \sqrt{d^2 - 4 n \sigma^2}} & \eqfor d^2 \geq 4 n \sigma^2, \\
 0 & \text{ else}.
\end{cases}
 \end{equation}
\end{prop}

Since the isotropic Gaussian matrix estimation problem has been thoroughly studied, we can compare the shrinkage rule $\psi(\cdot)$ with known asymptotically optimal ones. \citet{gavish2014optimalshrink} provide a comprehensive treatment of optimal singular-value shrinkage for different loss functions in a Marchenko-Pastur asymptotic regime, where $n$ and $p$ both diverge to infinity such that $p/n \rightarrow \beta$ for some $0 < \beta \leq 1$ while the rank and the scale of the signal remains fixed. This specific asymptotic setting has also been investigated by, among others, \citet{john2001asympeignull} and \citet{shabalin2013reconstruction}.

In what appears to be a remarkable coincidence, for the square case $\beta = 1$, our shrinkage rule \eqref{eq:iter_gauss} corresponds exactly to the Marchenko-Pastur optimal shrinkage rule under operator-norm loss $\Norm{\hmu - \mu}_{op}$; see the proof of Proposition \ref{prop:iter_gauss} for a derivation. At the very least, this connection is reassuring as it suggests that our iterative scheme may yield statistically reasonable estimates $\hmu^\iter$ for other noise models too. It remains to be seen whether this connection reflects a deeper theoretical phenomenon.

\section{Application: regularizing correspondence analysis}
\label{sec:ca}

When $X$ contains count data, we have a natural noise model $X_{ij} \sim \text{Poisson}(\mu_{ij})$ that is compatible with the L\'evy bootstrap, and so our stable autoencoding framework is easy to apply. In this situation, however, $X$ is often analyzed by correspondence analysis \citep{green1984ca,green2007ca} rather than using a direct singular-value decomposition. Correspondence analysis, a classical statistics technique pioneered by \citet{hirschfeld1935connection} and \citet{Benz69, Benz73}, underlies variants of many modern machine learning applications such as spectral clustering on graphs \citep[e.g.,][]{ng2002spectral,shi2000normalized} or topic modeling for text data (see Section \ref{sec:document}).
In this section, we show how to regularize correspondence analysis by stable autoencoding. This discussion also serves as a blueprint for extending our method to other low-rank multivariate techniques such as principal component analysis or canonical correlation analysis.

Correspondence analysis involves taking the singular-value decomposition of a transformed matrix $M$:
\begin{equation}
\label{eq:ca}
M = R^{-\frac12} \p{X - \frac{1}{N} rc^\top} C^{-\frac12}, \quad \where R = \diag\p{r}, \, C = \diag\p{c},
\end{equation}
$N$ is the the total number of counts, and $r$ and $c$ are vectors containing the row and column sums of $X$.
This transformation $M$ has several motivations. For example, suppose that $X$ is a 2--way contingency table, i.e., that we have $N$ samples for which we measure two discrete features $A \in \{1, \, ..., \, n\}$ and $B \in \{1, \, ..., \, p\}$, and $X_{ij}$ counts the number of samples with $A = i$ and $B = j$. Then $M$ measures the distance between $X$ and a hypothetical contingency table where $A$ and $B$ are independently generated with the same marginal distributions as before; in fact, the standard $\chi^2$--test for independence of $X$ uses $\Norm{M}_2^2$ as its test statistic.
Meanwhile, if $X$ is the adjacency matrix of a graph, then $M$ is a version of the symmetric normalized graph Laplacian where we have projected out the first trivial eigencomponent.
Once we have a rank-$k$ estimate of $\hM_k$ obtained as in \eqref{eq:svd_mu}, we get
\begin{equation}
\label{eq:ca_restore}
\hmu_k^{CA} = R^{\frac12} \,  \hM_k  \,C^{\frac12} + \frac{1}{N} rc^\top.
\end{equation}
Our goal is to get a better estimator $\hM$ for the matrix $M$ of the population; we then transform $\hM$ into an estimate of $\mu$ using the same formula \eqref{eq:ca_restore}.

Following \eqref{eq:autoencoder_boot}, we propose regularizing the choice of $M$ as follows:
\begin{align}
\label{eq:autoencoder_ca}
&\hM_k^\stable = M\hB_k, \where \\
\notag
&\hB_k = \argmin_B \left\{\EE[\tX \sim \tlaw_\delta(X)]{\Norm{M - R^{-\frac12} \p{\tX - \frac{1}{N} rc^\top} C^{-\frac12} B}_2^2} : \rank{B} \leq k \right\}.
\end{align}
Just as in Theorem \ref{theo:compute}, we can show that $\hB_k$ solves
\begin{equation}
\label{eq:ca_opt}
\hB_k = \argmin_{B} \left\{ \Norm{M - MB}_2^2 + \Norm{S_M^{\frac12} B} : \rank{B} \leq k  \right\},
\end{equation}
where $S_M$ is a diagonal matrix with
$\p{S_M}_{jj} = {c_j^{-1}} \sum_{i = 1}^n  \text{Var}_{\tX \sim \tlaw_\delta(X)}[\tX_{ij}] / {r_i}$.
We can efficiently solve for \eqref{eq:ca_opt} using the same method as in \eqref{eq:nongauss_svd} and \eqref{eq:closed_form}. Finally, if we do not want to fix the rank $k$, we can use an iterative scheme as in Section \ref{sec:iter}.

Since $X$ contains count data, we generate the bootstrap samples $\tX \sim \tlaw_\delta(X)$ using the Poisson-compatible bootstrap algorithm \eqref{eq:poisson_noise}, i.e., $\tX_{ij} \sim (1 - \delta)^{-1} \operatorname{Binomial}\p{X_{ij}, \, 1 - \delta}$.
Interestingly, if we had chosen to sample $\tX$ from an independent contingency table with
\begin{equation}
\EE[\tX \sim \tlaw_\delta]{\tX} = \frac{1}{N} rc^\top, \, \Var[\tX \sim \tlaw_\delta]{\tX_{ij}} = \frac{\delta}{1 - \delta} \frac{r_ic_j}{N},
\end{equation}
we would have obtained a regularization matrix $S_M = n \delta / (N (1 - \delta)) I_{p \times p}$. Because $S_M$ is diagonal, the resulting estimator $\hM_\lambda$ could then be obtained from $M$ by singular value shrinkage. Thus, if we want to regularize correspondence analysis applied to a nearly independent table, singular value shrinkage based methods can achieve good performance; however, if the table has strong dependence, our framework provides a more principled way of being robust to sampling noise.

\section{Simulation experiments}

To assess our proposed methods, we first run comparative simulation studies for different noise models. We begin with a sanity check: in Section \ref{sec:gauss}, we reproduce the isotropic Gaussian noise experiments of \citet{candes2013sure}, and find that our method is competitive with existing approaches on this standard benchmark.

We then move to the non-isotropic case, where we can take advantage of our method's ability to adapt to different noise structures. In Section \ref{sec:poissonnoise} we show results on experiments with Poisson noise, and find that our method substantially outperforms its competitors. Finally, in Section \ref{sec:realworld}, we apply our method to real-world applications motivated by topic modeling and sensory analysis.

\subsection{Gaussian noise}
\label{sec:gauss}

We compare our estimators to existing ones by reproducing the simulations of \citet{candes2013sure}. For this experiment, we generated data matrices of size  $200\times 500$  according to the Gaussian noise model \eqref{eq:gauss} with four signal-to-noise ratios SNR$\in \{0.5, 1, 2, 4\}$ calculated as $1/(\sigma\sqrt{np})$, and two values for the underlying rank $k \in \{10, 100\}$; results are in Table \ref{simu:candes}.

\paragraph{Methods under consideration:}
Our goal is to evaluate the performance of the stable autoencoder (SA) as defined in \eqref{eq:autoencoder_boot} and the iterated stable autoencoder (ISA) described in Algorithm \ref{alg:iter}. As discussed in Section \ref{sec:boot}, we applied our stable autoencoding methods to $X^\top$ rather than $X$, so that $n$ was larger than $p$; and set the tuning parameter to $\delta = 1/2$. For ISA, we ran the iterative Algorithm \ref{alg:iter} for 100 steps, although the algorithm appeared to become stable after 10 steps already.

In addition to our two methods, we also consider the following estimators:
\begin{itemize}
 \item Truncated SVD with fixed rank $k$ (TSVD-$k$). This is the classical approach  \eqref{eq:svd_mu}.
 \item Adaptively truncated SVD (TSVD-$\tau$), using the asymptotically optimal threshold of \citet{gavish2014optimal}.
 \item Asymptotically optimal singular-value shrinkage (ASYMP) in the Marchenko-Pastur asymptotic regime given the Frobenius norm loss \citep{shabalin2013reconstruction,gavish2014optimalshrink}, with shrinkage function
\begin{equation}
\label{eq:shaba}
\psi(d) =
\begin{cases}
\frac{1}{d} \sqrt{\left(d^2 - \p{1 + \beta} n \sigma^2\right)^2 - 4 \beta n^2 \sigma^4} & \eqfor d^2 \geq \p{1 + \sqrt{\beta}}^2 n \sigma^2, \\
0 & \text{ else,}
\end{cases}
\end{equation}
where $\beta = p/n$ is the aspect ratio, assuming without loss of generality that $p \leq n$.
 \item The shrinkage scheme of \citet{verbanck2013regularised} motivated by low-noise (LN) asymptotics.  It uses  for $\psi\p{d_l}$ in \eqref{eq:svd_shrink}, 
 \begin{equation}\label{eq:twosteps}
\psi\p{d_l}= \begin{cases}
d_l \left(1-\frac{\sigma^2}{d_l^2}\right)  & \eqfor l \leq k, \\
0 & \text{ else.}
\end{cases}
\end{equation}
 \item Singular value soft thresholding (SVST) \citep{cai2010singular}, where the singular values are soft-thresholded by $\tau$ selected by minimizing a Stein unbiased risk estimate (SURE), as suggested by \citet{candes2013sure}.
\end{itemize}
All the estimators are defined assuming the variance of the noise scale $\sigma^2$ to be known.  In addition, TSVD-$k$, SA, and LN require  the rank $k$ as a tuning parameter. In this case, we set $k$ to the true rank of the underlying signal.

%We report in Table~\ref{simu:candes} the estimated mean squared error between the fitted matrices $\hmu$ obtained from the different methods and the true signal $\mu$ as well as the estimated rank (the number of singular values that are not set to zero).

\spacingset{\SPACESMALL}
\begin{table}
\begin{center}
\begin{tabular}{rr||c|c|c|c|c|c|c|}
\multicolumn{1}{r}{$k$}	&	\multicolumn{1}{c||}{SNR}	&
\multicolumn{2}{c|}{Stable} & 
\multicolumn{2}{c|}{TSVD} & \multicolumn{1}{c|}{ASYMP} & \multicolumn{1}{c|}{SVST} & \multicolumn{1}{c|}{LN}  \\ \hline
        &               &  
        \multicolumn{1}{c}{${\rm SA}$}  & \multicolumn{1}{c|}{${\rm ISA}$} &        
        \multicolumn{1}{c}{$k$} & \multicolumn{1}{c|}{$\tau$} & \multicolumn{1}{c|}{} & \multicolumn{1}{c|}{}     &  \multicolumn{1}{c|}{} 
           \\ \hline
        {\bf MSE} &&&&&&&& \\
10&4 & \textbf{0.004}&\textbf{0.004}&
\textbf{0.004} & \textbf{0.004} &  \textbf{0.004} &0.008&\textbf{0.004}\\
100&4&\textbf{0.037}&\textbf{0.036}&
\textbf{0.038} &\textbf{0.038} &  \textbf{0.037}   & 0.045&\textbf{0.037}\\
10&2&\textbf{0.017}&\textbf{0.017}&
\textbf{0.017} & \textbf{0.016} &   \textbf{0.017}     & 0.033&\textbf{0.017}\\
100&2 &{\bf 0.142}&\textbf{0.143}&
0.152 & 0.158  &  \textbf{0.146}    & 0.156&\textbf{0.141}\\
10&1& \textbf{0.067}&\textbf{0.067} &
0.072 & 0.072 &  \textbf{0.067}     & 0.116&\textbf{0.067}\\
100&1&0.511&0.775&
0.733 &0.856 &  0.600    & \textbf{0.448}&0.491\\
10&0.5&0.277& \textbf{0.251}&
0.321 & 0.321 &  \textbf{0.250}   & 0.353&0.257\\
100&0.5&1.600&1.000&
3.164 &1.000 &  0.961    &\textbf{0.852}&1.477\\\hline
{\bf Rank}&&&&&&&& \\
10&4&   & \textbf{10}  &   &  \textbf{10} & \textbf{10}    & 65 &   \\
100&4&  &  \textbf{100} &  & \textbf{100} & \textbf{100}     & 193 &  \\
10&2&    &  \textbf{10} &  & \textbf{10}  & \textbf{10}     & 63 &  \\
100&2 & & \textbf{100} &  & \textbf{100}  & \textbf{100}     & 181 &  \\
10&1&    &  \textbf{10} &   & \textbf{10}  & \textbf{10}    & 59 &    \\
100&1 & &  29.6 &  & 38 &  \textbf{64}    & 154 &  \\
10&0.5&   & \textbf{10}  &   & \textbf{10} &  \textbf{10}    & 51 &   \\
100&0.5 &  &   0 & &   0 &  15     &  \textbf{86} &  \\ \hline
\end{tabular}
\caption{Mean cell-wise squared error (top) and rank estimates (bottom) obtained by  the methods described in Section \ref{sec:gauss}, averaged over 50 simulation replications.
The best results for each row are indicated in {bold}.
\label{simu:candes}}
\end{center}
 \end{table}
 \spacingset{\SPACEBIG}

As our simulation study makes clear, the proposed methods have very different strengths and weaknesses. 
Both methods that apply a hard thresholding rule to the singular values, namely TSVD-$k$ and TSVD-$\tau$, provide accurate MSE when the SNR is high but break down in low SNR settings. 
Conversely, the SVST behaves well in low SNR settings, but struggles in other regimes. 
This is not surprising, as the method over-estimates the rank of $\mu$. This behavior is reminiscent of what happens in lasso regression \citep{tibshirani1996regression} when too many variables are selected \citep{zou2006adapt,zhang2008lasso}.

Meanwhile, the estimators with non-linear singular-value shrinkage functions, namely SA, ISA, ASYMP, and LN are more flexible and perform  well except in the very difficult scenario where the signal is overwhelmed by the noise. 
Both ASYMP and ISA estimate the rank accurately except when the signal is nearly indistinguishable from the noise (SNR=0.5 and $k$=100).

\subsection{Poisson noise}
\label{sec:poissonnoise}

\begin{figure}
\begin{center}
\includegraphics[width = 0.3\textwidth]{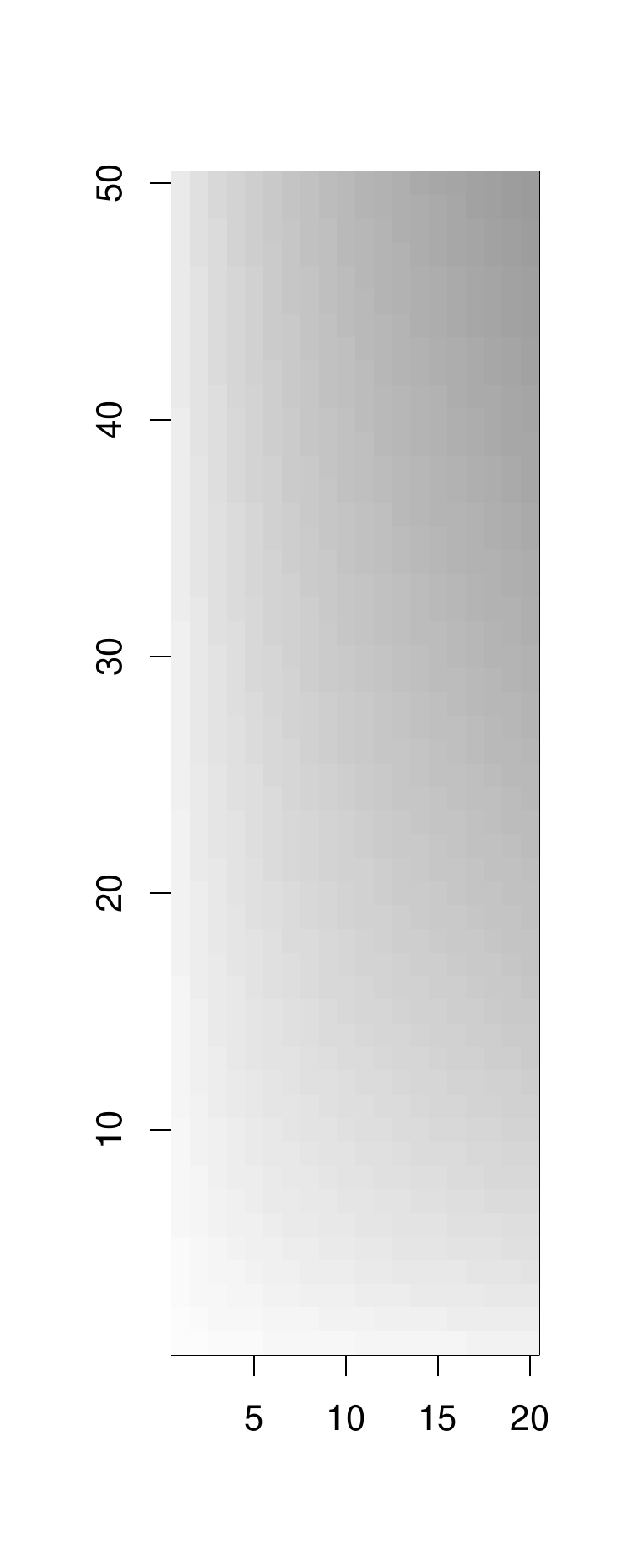}
\hspace{0.02\textwidth}
\includegraphics[width = 0.3\textwidth]{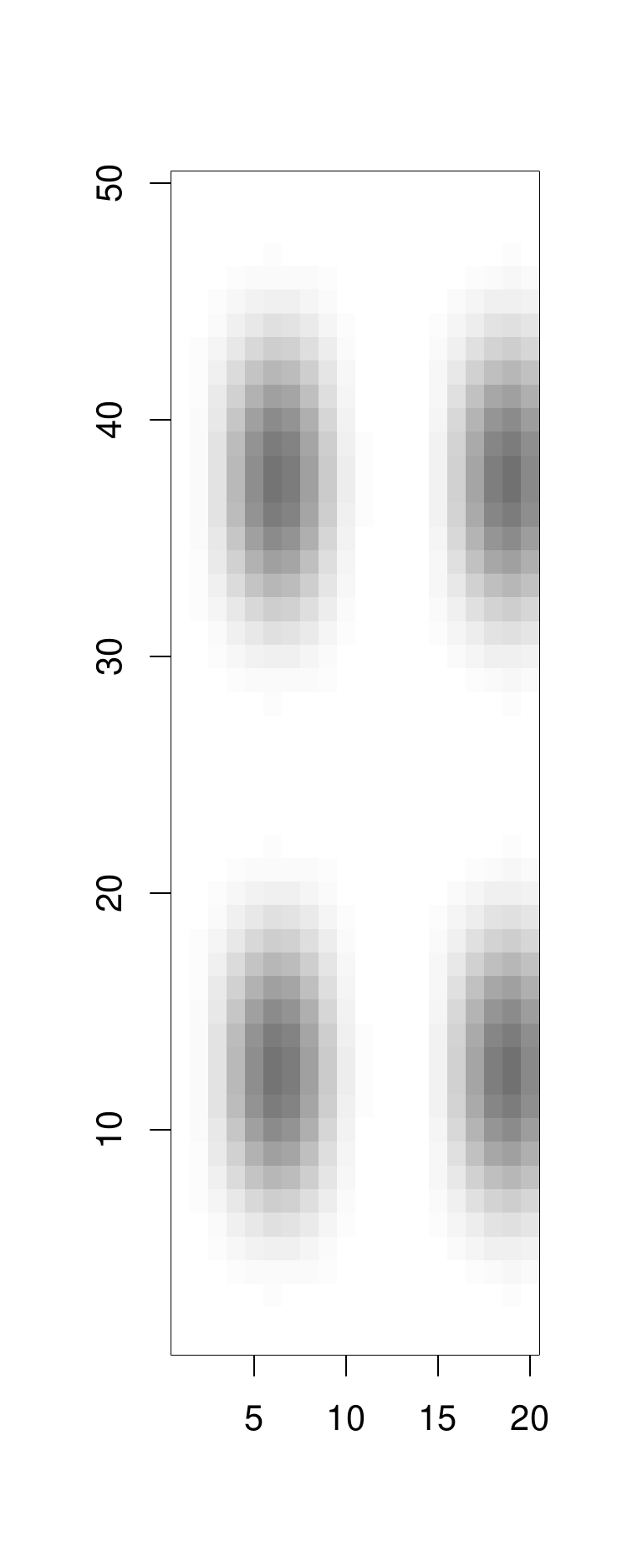}
\hspace{0.02\textwidth}
\includegraphics[width = 0.3\textwidth]{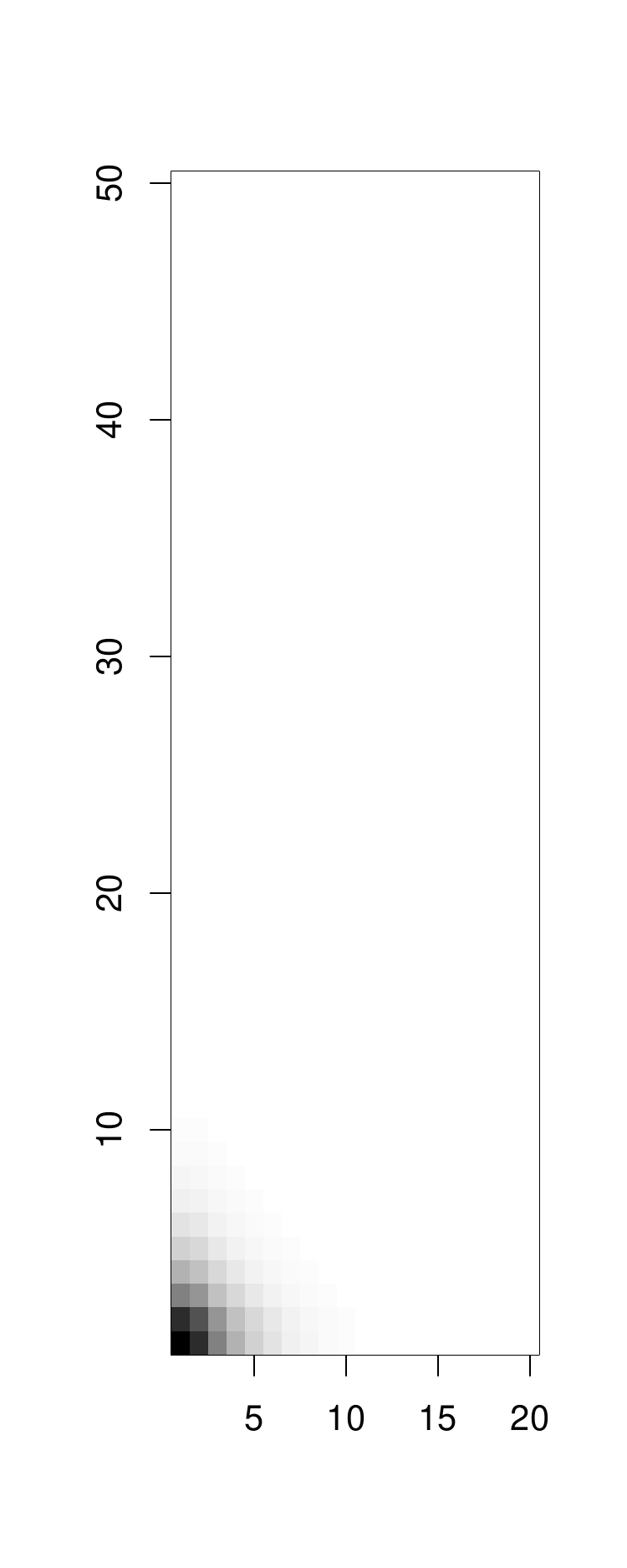}
\caption{The 3 components of the mean of the underlying Poisson process; the dark areas have the highest intensity. The corresponding singular values have relative magnitudes $1.1:1.4:1$.}
\label{fig:mu}
\end{center}
\end{figure}

Once we move beyond the isotropic Gaussian case, our method can both learn better singular vectors and out-perform its competitors in terms of MSE. We illustrate this phenomenon with a simple simulation example, where we drew $X$ of size $n=50$ and $p=20$ from a Poisson distribution with expectation $\mu$ of rank 3 represented in Figure \ref{fig:mu}. Because the three components of $\mu$ have different levels of concentration---the first component is rather diffuse, while the third one is concentrated in a corner---adapting to the Poisson variance structure is important.

We varied the effective signal-to-noise ratio by adjusting the mean number of counts in $X$, i.e., $N = \sum_{ij} \mu_{ij}$. We then report results  for the normalized mean matrix $\mu/N$.
We used both SA and ISA to estimate $\mu$ from $X$; in both cases, we generated $\tX$ with the Poisson-compatible bootstrap noise model \eqref{eq:poisson_noise}, and set $\delta = 1/2$.
We also used LN, ASYMP and TSVD-$\tau$ as baselines, although they are only formally motivated in the Gaussian model. These methods require a value for $\sigma$. For LN, we used the method recommended by \citet{josse2011gcvpca}:
\begin{equation}
\label{eq:sigma_hat}
\hat \sigma^2= \frac{\Norm{X- \sum_{l=1}^ku_ld_lv_l}_2^2 }{np-nk-kp+k^2}.
\end{equation}
For ASYMP and TSVD-$\tau$, we used the estimator suggested in \citet{gavish2014optimalshrink},
%\begin{equation}
%\label{eq:sigma_hatdono}
$\hat \sigma= {d_{med}} \, / \, {\sqrt{n \mu_{\beta}}}$,
%\end{equation}
where $d_{med}$ is the median of the singular values of $X$ and $\mu_{\beta}$ is the median of the Marcenko-Pastur distribution with aspect ratio $\beta$.

\spacingset{\SPACESMALL}
 \spacingset{\SPACEBIG}

\spacingset{\SPACESMALL}
\begin{table}[t]
\begin{center}
\begin{tabular}{r||c|c|c|c|c|c|}
\multicolumn{1}{r||}{$N$}	&
\multicolumn{2}{c|}{Stable} & 
\multicolumn{2}{c|}{TSVD} & \multicolumn{1}{c|}{ASYMP} & \multicolumn{1}{c|}{LN}  \\ \hline
     
     &   \multicolumn{1}{c}{${\rm SA}$}  & \multicolumn{1}{c|}{${\rm ISA}$} &        
        \multicolumn{1}{c}{$k$} & \multicolumn{1}{c|}{$\tau$} & \multicolumn{1}{c|}{}    &  \multicolumn{1}{c|}{} \\\hline
  \hline
200 & 1.83 & \bf 1.13 & 2.62 & 1.99 & 1.71 & 2.12 \\ 
  400 & 0.76 & \bf 0.51 & 1.08 & 0.93 & 0.77 & 0.88 \\ 
  600 & 0.46 & \bf 0.36 & 0.63 & 0.58 & 0.48 & 0.52 \\ 
  800 & 0.33 & \bf 0.29 & 0.44 & 0.42 & 0.35 & 0.37 \\ 
  1000 & \bf 0.25 & \bf 0.24 & 0.32 & 0.33 & 0.27 & 0.28 \\ 
  1200 & \bf 0.20 & \bf 0.19 & 0.25 & 0.27 & 0.22 & 0.22 \\ 
  1400 & \bf 0.16 & \bf 0.15 & 0.20 & 0.22 & 0.19 & 0.18 \\ 
  1600 & \bf 0.14 & \bf 0.13 & 0.17 & 0.19 & 0.16 & 0.15 \\ 
  1800 & \bf 0.12 & \bf 0.11 & 0.14 & 0.16 & 0.14 & 0.13 \\ 
  2000 & \bf 0.11 & \bf 0.10 & 0.13 & 0.15 & 0.13 & 0.12 \\ 
   \hline

 \hline
\end{tabular}
\caption{Mean cell-wise squared error, averaged over 1000 simulation replications.  
\label{simu:poisson_mse}}
\end{center}
 \end{table}
 \spacingset{\SPACEBIG}

\spacingset{\SPACESMALL}
\begin{table}
\begin{center}
\begin{tabular}{r||c|c|c|c|c|c|}
\multicolumn{1}{r||}{$N$}	&
\multicolumn{3}{c|}{RV for $U$} & 
\multicolumn{3}{c|}{RV for $V$}   \\ \hline
     
     &   \multicolumn{1}{c}{${\rm SVD}$}  & \multicolumn{1}{c}{${\rm SA}$} & \multicolumn{1}{c|}{${\rm ISA}$} 
      &   \multicolumn{1}{c}{${\rm SVD}$}  & \multicolumn{1}{c}{${\rm SA}$} & \multicolumn{1}{c|}{${\rm ISA}$}  \\\hline
  \hline
200 & 0.29 & \bf 0.34 & - & 0.34 & \bf 0.40 & - \\ 
  400 & 0.48 & \bf 0.53 & 0.51 & 0.53 & \bf 0.57 & 0.54 \\ 
  600 & 0.60 & 0.64 & \bf 0.71 & 0.64 & 0.69 & \bf 0.79 \\ 
  800 & 0.67 & 0.71 &\bf  0.79 & 0.72 & 0.76 & \bf 0.87 \\ 
  1000 & 0.74 & 0.77 & \bf 0.82 & 0.79 & 0.83 & \bf 0.89 \\ 
  1200 & 0.78 & 0.81 & \bf 0.85 & 0.84 & 0.87 & \bf 0.90 \\ 
  1400 & 0.82 & \bf 0.85 & \bf 0.86 & 0.87 & 0.90 & \bf 0.92 \\ 
  1600 & 0.85 & \bf 0.87 & \bf 0.88 & 0.90 & 0.91 & \bf 0.93 \\ 
  1800 & 0.87 & \bf 0.88 & \bf 0.89 & 0.92 & \bf 0.93 & \bf 0.94 \\ 
  2000 & 0.88 & \bf 0.89 & \bf 0.90 & \bf 0.93 & \bf 0.94 & \bf 0.94 \\ 
   \hline
\end{tabular}
\caption{RV coefficients between the estimated and true $U$ and $V$ matrices, averaged over 1000 simulation replications. For ISA, we only averaged performance over examples where the estimated rank was at least 3. In the $N=200$  case, no results is given for ISA since the estimated rank is always less than 3. 
\label{simu:poisson_rv}}
\end{center}
 \end{table}
 \spacingset{\SPACEBIG}

\spacingset{\SPACESMALL}
\begin{table}
\begin{center}
\begin{tabular}{r||c|c|c|}
\multicolumn{1}{r||}{$N$}	&
 \multicolumn{1}{c}{${\rm TSVD-}\tau$}  & \multicolumn{1}{c}{${\rm ASYMP}$} & \multicolumn{1}{c|}{${\rm ISA}$}  \\\hline
  \hline
200 & 1.78 & 3.11 & 1.40 \\ 
  400 & 2.23 & 3.55 & 1.96 \\ 
  600 & 2.54 & 3.77 & 2.01 \\ 
  800 & 2.76 & 3.9 & 2.10 \\ 
  1000 & 2.94 & 3.99 & 2.36 \\ 
  1200 & 3.11 & 4.02 & 2.71 \\ 
  1400 & 3.17 & 4.04 & 2.92 \\ 
  1600 & 3.17 & 4.06 & 2.98 \\ 
  1800 & 3.22 & 4.08 & 3.00 \\ 
  2000 & 3.23 & 4.07 & 3.00 \\ 
   \hline
\end{tabular}
\caption{Mean rank estimates for the Poisson simulation, averaged over 1000 simulation replications. The true rank of the underlying signal is 3.
\label{simu:poisson_k}}
\end{center}
 \end{table}
 \spacingset{\SPACEBIG}

In addition to providing MSE (Table \ref{simu:poisson_mse}), we also report the alignment of the row/column directions $U$ and $V$ with those of the true mean matrix $\mu$  (Table \ref{simu:poisson_rv}). We measured alignment using the RV coefficient, which is a matrix version of Pearson's squared correlation coefficient $\rho^2$ that takes values between 0 and 1 (\citet{escoufier1973RV}; see \citet{josse2013measures} for a review):
\begin{equation}
\mbox{RV}(U, \, \hat U) = {\mbox{tr}\p{U^{\top}\hat U \hat U^{\top}U}} \ \Bigg/ \ {\sqrt{\mbox{tr}\p{\p{U^{\top}U}^2}\mbox{tr}\p{\p{\hat U^{\top} \hat U}^2}}}.
\end{equation}
Finally, we also report the mean estimated ranks in Table \ref{simu:poisson_k}.

We see that our methods based on stable autoencoding do well across all noise levels. In the high-noise setting (i.e., with a small number of count observations $N$), the iterated stable autoencoder does particularly well, as it is able to use a lower rank in response to the weaker signal. As seen in Table \ref{simu:poisson_rv}, the ability to learn new singular vectors appears to have been useful here, as the ``$\widehat{U}$'' and ``$\widehat{V}$'' matrices obtained by stable autoencoding are much better aligned with the population ones than those produced by the SVD are. We also see that, in the low noise setting where $N$ is large, ISA recovers the true rank $k = 3$ almost exactly, whereas ASYMP and TSVD-$\tau$ do not.
Finally, we note that all shrinkage methods did better than the baseline, namely the simple rank-3 SVD. Thus, even though LN, ASYMP and TSVD-$\tau$ are only formally motivated in the Gaussian noise case, our results suggest that they are still better than no regularization on generic problems.

\section{Real-world examples}
\label{sec:realworld}

To highlight the wide applicability of our method, we use it on two real-world problems from different fields. We begin with a larger natural language application, where we use the iterated stable autoencoder to improve the quality of topics learned by latent semantic analysis, and evaluate results by end-to-end classifier performance. Next, in Section \ref{sec:perfume}, we analyze a smaller dataset from a consumer survey, and show how our regularization schemes can improve the faithfulness of correspondence analysis graphical outputs commonly used by statisticians.

\subsection{Learning topics for sentiment analysis}
\label{sec:document}

\begin{table}[t]
\centering
\begin{tabular}{r|ccc|}
 & Document Averaged & Corresp. Analysis & Corresp. Analysis + ISA \\ \hline
Accuracy & 62.1 \% & 61.8 \% & {\bf 67.0} \% \\
Times Best & $2/10,000$ & $1/10,000$ & $\mathbf{9,997} / 10,000$ \\ \hline
\end{tabular}
\caption{Test set accuracy of a logistic regression classifier trained on topics learned by latent semantic analysis, averaged over 10,000 train/test splits. The topic models were run only once on all the (unlabeled) data; thus we are in a transductive setting. Each method used $k = 5$ topics; this number was automatically picked by ISA.}
\label{tab:rt2k}
\end{table}

Many tasks in natural language processing involve computing a low-rank approximation
to a \emph{document/term-frequency} matrix $X$, i.e., $X_{ij}$ counts the number of times
word $j$ appears in document $i$. The singular rows of $X$ can then be interpreted as
\emph{topics} characterized by the prevalence of different words, and each document is described as a
mixture of topics. The idea of learning topics using an SVD of (a normalized version of) the
matrix $X$ is called latent semantic analysis \citep{deerwester1990indexing}.
Here, we argue that we can make the topics discovered by latent semantic analysis
better by regularizing the SVD of $X$ using an iterated stable autoencoder.

To do so, we examine the Rotten Tomatoes movie review dataset collected by \citet{pang2004sentimental},
with $n = 2,000$ documents and $p = 50,921$ unique words. We learned topics with three variants
of latent semantic analysis, which involve using different transformations/regularization schemes
while taking an SVD.
\begin{itemize}
\item Document averaging: in order to avoid large documents dominating the fit, we compute the matrix $\Pi_{ij} = X_{ij} / \sum_j X_{ij}$. We then perform a rank-$k$ SVD of $\Pi$.
\item Correspondence analysis: we run a rank-$k$ SVD on $M$ from \eqref{eq:ca}. This approach  normalizes by both \smash{$R^{-\frac12}$ and $C^{-\frac12}$} in order to counteract the excess influence of both long documents and common words, instead of just using $\Pi = R^{-1} X$ as above.
\item ISA-regularized correspondence analysis ($\delta = 0.5$).
\end{itemize}
Correspondence analysis with ISA picked $k = 5$ topics; we also used $k = 5$ for the other methods. The document averaging method did not appear to benefit much from regularization; presumably, this is because the matrix $\Pi$ does not up-weight rare words. 

Because $n$ and $p$ are both fairly large, it is difficult to evaluate the quality of the learned topics directly. To get around this, we used a more indirect approach and examined the quality of the learned decompositions of $X$ by using them for sentiment classification. Specifically, each method produces a low-rank decomposition $UDV^\top$, where $U$ is an $n \times k$ orthonormal matrix; we then used the columns of $U$ as features in a logistic regression. We trained the logistic regression on one half of the data and then tested it on the other half, repeating this process over 10,000 random splits. We are in a transductive setting because we used all the data (but not the labels) for learning the topics.

The results, shown in Table \ref{tab:rt2k}, suggests that ISA substantially improves the performance of latent semantic analysis for this dataset. In a somewhat surprising twist, we may have expected the decomposition based on correspondence analysis to out-perform the baseline that just divides by document length; however, correspondence analysis ended up doing slightly worse. The problem appears to have been that, because correspondence analysis up-weights less common words relative to the common ones, its topics become more vulnerable to noise. Thus, it is not able to beat document-wise averaging although it has a seemingly better normalization scheme. But, once we use ISA to regularize it, correspondence analysis is able to fully take advantage of the down-weighting of common words.

\begin{figure}[t]
\centering
\includegraphics[width=0.45\textwidth]{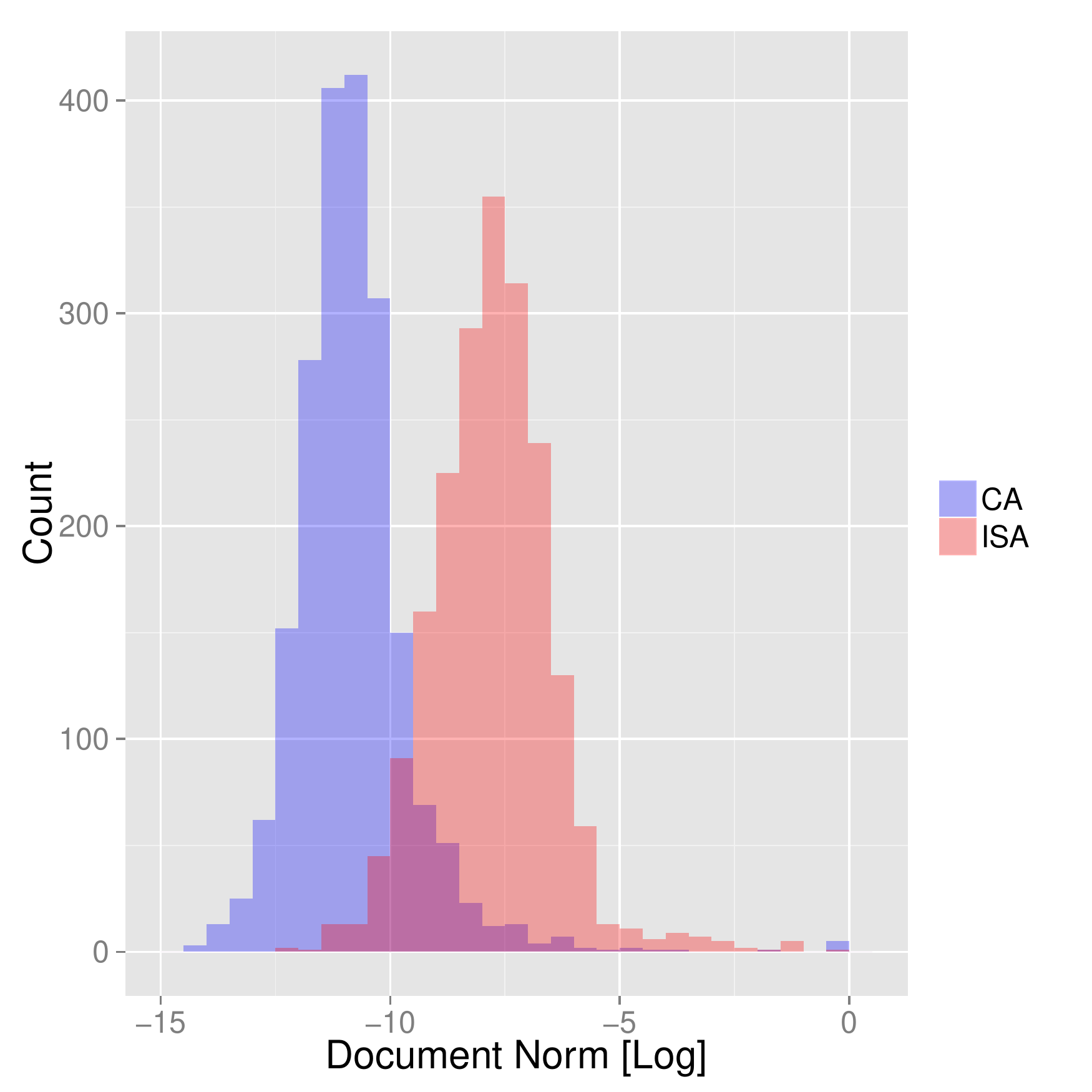}
\caption{Distribution of $\log \Norm{U_{i\cdot}}_2^2$ for correspondence analysis with and without ISA. Using ISA increases the influence of the median document in learning topics.}
\label{fig:rt2k}
\end{figure}

We can visualize the effect of regularization using Figure \ref{fig:rt2k}, which shows the distribution of $\log \Norm{U_{i\cdot}}_2^2$ for the $U$-matrices produced by correspondence analysis with and without ISA. The quantity $\Norm{U_{i\cdot}}_2^2$ measures the importance of the $i$-th document in learning the topics. We see that plain correspondence analysis has some documents that dominate the resulting $U$ matrix, whereas with ISA the magnitudes of the contributions of different documents are more evenly spread out. Thus, assuming that we do not want topics to be dominated by just a few documents, Figure \ref{fig:rt2k} corroborates our intuition that ISA improves the topics learned by correspondence analysis.

Finally, we note that there exist several topic models that do not reduce to an SVD \citep[e.g.,][]{blei2003latent,hofmann2001unsupervised,xu2003document}; in fact, one of the motivations for latent Dirichlet allocation \citep{blei2003latent} was to add regularization to topic modeling using a hierarchical Bayesian approach.
Moreover, methods that only rely on unsupervised topic learning do not in general achieve state-of-the-art accuracy for sentiment classification on their own \citep[e.g.,][]{wang2012baselines}. Thus, our goal here is not to advocate an end-to-end methodology for sentiment classification, but only to show that stable autoencoding can substantially improve the quality of topics learned by a simple SVD in cases where a practitioner may want to use them.

\subsection{A sensory analysis of perfumes}
\label{sec:perfume}

Finally, we use stable autoencoding to regularize a sensory analysis of perfumes. The data for the analysis was collected by asking consumers to describe 12 luxury perfumes such as \textit{Chanel Number 5} and \textit{J'adore} with words. The answers were then organized in a  $12 \times 39$ (39 words unique were used) data matrix where each cell represents the number of times a word is associated to a perfume; a total of $N = 1075$ were used overall. The dataset is available at \texttt{http://factominer.free.fr/docs/perfume.txt}.
We used correspondence analysis (CA)  to visualize the associations between words and perfumes.
Here, the technique allows to highlight perfumes that were described using a similar profile of words, and to find words that describe the differences between groups of perfumes.

In order to get a better idea of which regularization method is the most trustworthy here, we ran a small bootstrap simulation study built on top of the perfume dataset. We used the full $N = 1075$ perfume dataset as the population dataset, and then generated samples of size $N=200$ by subsampling the original dataset without replacement. Then, on each sample, we performed a classical correspondence analysis by performing a rank-$k$ truncated SVD of the matrix $M$ \eqref{eq:ca}, as well as several regularized alternatives described in Section \ref{sec:gauss}.

For each estimator, we report  its singular values, as well as the RV-coefficients between its row (respectively column) coordinates and the population ones. All the methods except for ISA require us to specify the rank $k$ as an input parameter. Here, of course, $k$ is unknown since we are working with a real dataset; however, examining the full-population dataset suggests that using $k = 2$ components is appropriate.
For LN, SA and ISA, we set tuning parameters as in Section \ref{sec:poissonnoise}, namely LN uses $\hsigma$ from \eqref{eq:sigma_hat}, while SA is performed with $\delta=0.5$.  For ISA we used $\delta = 0.3$; this latter choice was made to get rank-2 estimates.  In practice, one could also consider cross-validation to find a good value for $\delta$.

\spacingset{\SPACESMALL}
\begin{table}[t]
\centering
\centering
\begin{tabular}{r|ccccc|}
  \hline
 & $d_1$ & $d_2$ & RV row & RV col & $k$\\ 
  \hline
TRUE & 0.44 & 0.15 & & &\\
CA & 0.62 & 0.42 &   0.41 & 0.72& \\ 
  LN & 0.28 & 0.11 &  0.47 & 0.79& \\ 
   SA & 0.34 & 0.18 &   0.50 & 0.79& \\ 
 ISA & 0.40 & 0.18  & \bf 0.52 & \bf 0.81 & 2.43\\ 
   \hline
\end{tabular}
\caption{Performance of standard correspondence analysis (CA) as well as regularized alternatives on the perfume dataset. We report singular values, RV-coefficients and rank estimates; results correspond to the mean over the 1000 simulations. 
\label{tab:perfume}
}
\end{table}
\spacingset{\SPACEBIG}

\spacingset{\SPACESMALL}
\begin{figure}[p]
    \begin{center}
   \includegraphics[width=0.9\textwidth]{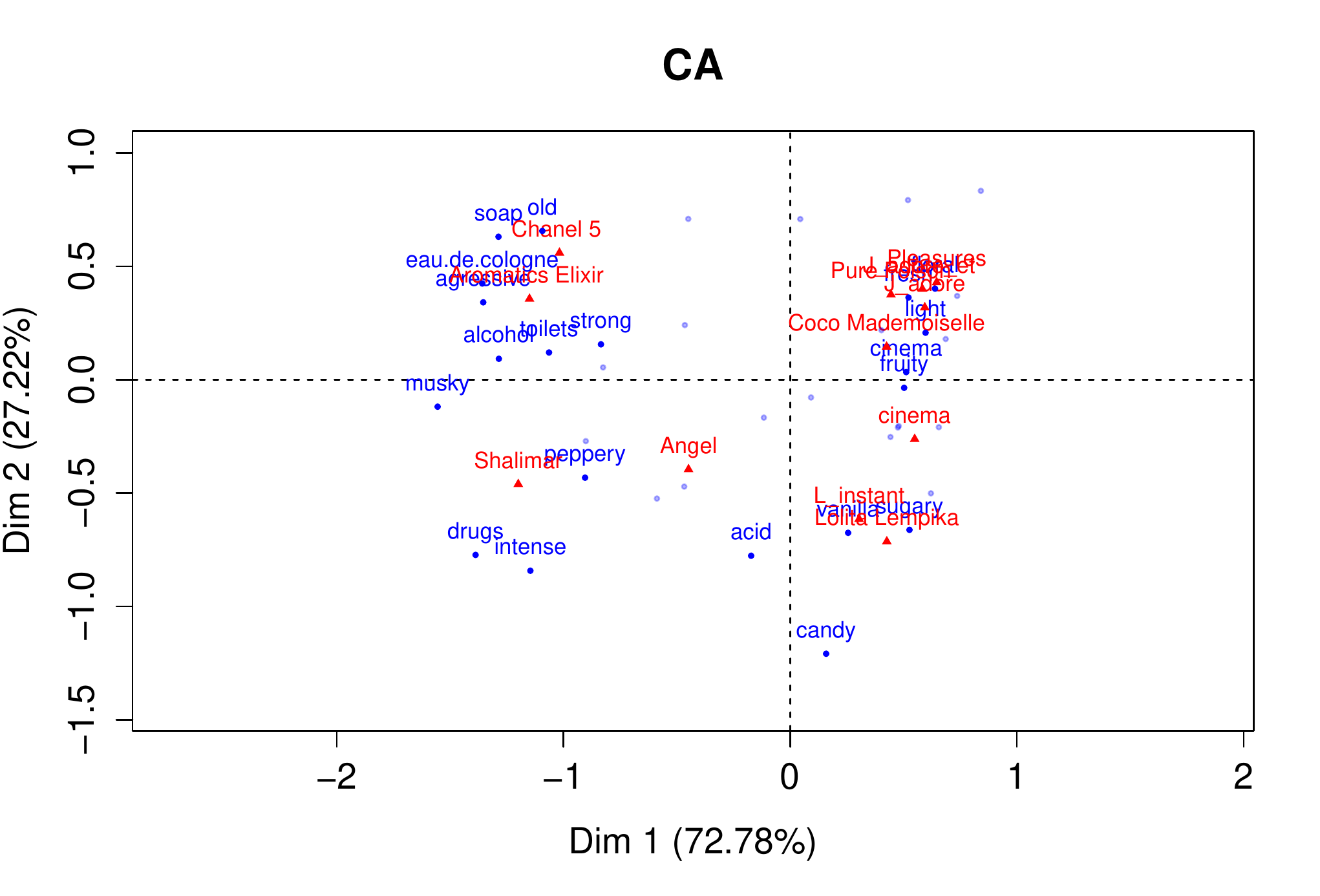} 
\includegraphics[width=0.9\textwidth]{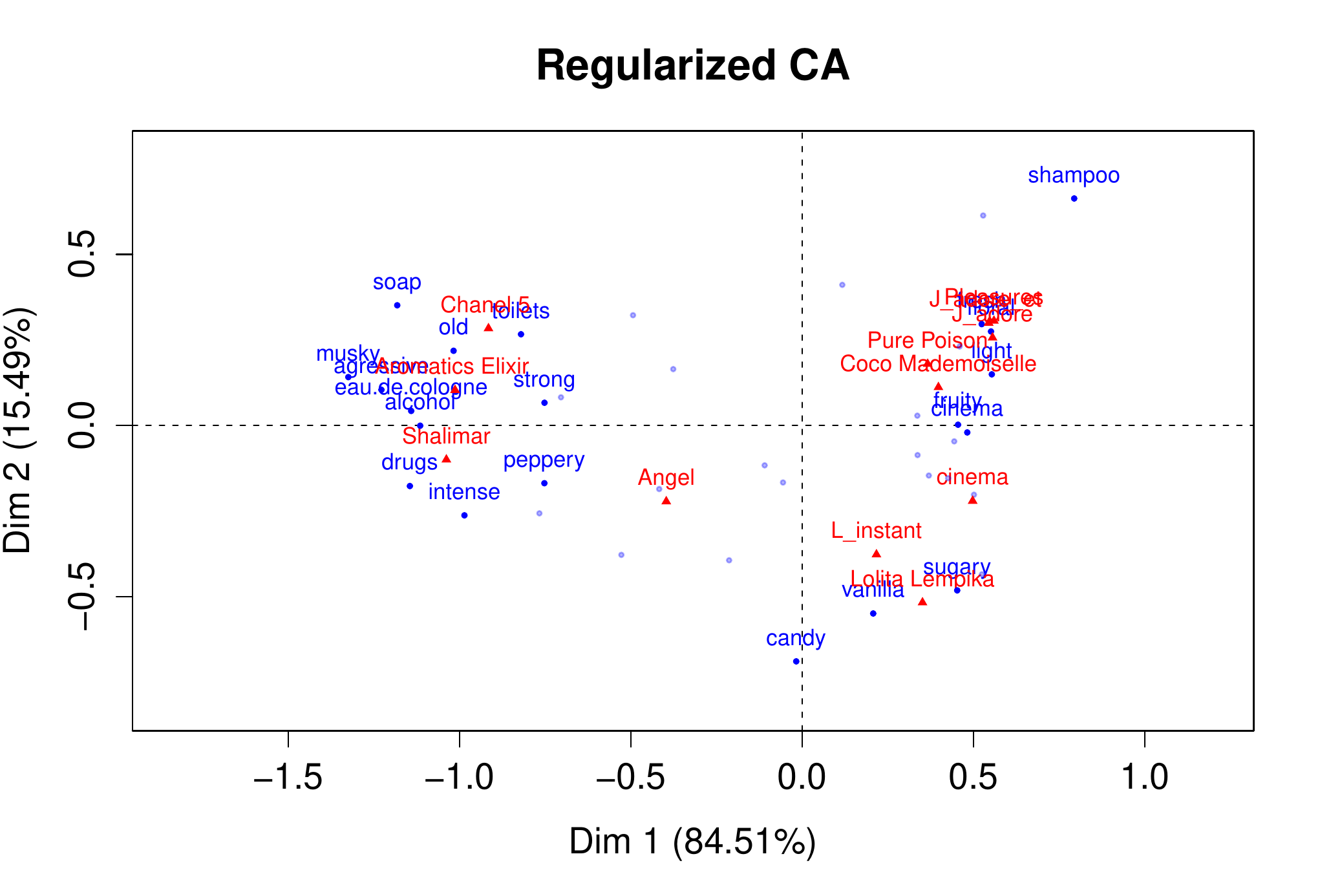}
 \caption{Results for CA on a sample data set (top) and Regularized CA (bottom) using ISA on a single subsample of size $N = 200$. Only the 20 words that contribute the most to the dimensions of variability are represented.}
\label{fig:perfume}
    \end{center} 
  \end{figure}
\spacingset{\SPACEBIG}

Results are shown in Table \ref{tab:perfume}. From a practical point of view, it is also interesting to compare the graphical output of correspondence analysis with and without regularization. Figure \ref{fig:perfume} (top) shows two-dimensional CA representation on one sample and Figure \ref{fig:perfume} (bottom) shows the representation obtained with ISA. Only the 20 words that contribute the most to the first two dimensions are represented.  The analysis is performed using the R package FactoMineR \citep{hussonfacto2008}.

Our results emphasize that, although correspondence analysis is often used as a visualization technique, appropriate regularization is still important, as regularization may substantially affect the graphical output. For example, on the basis of the CA plot, the perfume {\it Shalimar} looks like an outlier, whereas after regularization it seems to fit in a cluster with {\it Chanel 5} and {\it Elixir}. We know from Table \ref{tab:perfume} that the regularized CA plots are better aligned with the population ones than the unregularized ones are; thus, we may be more inclined to trust insights from the regularized analysis.
 
\section{Discussion}

In this paper, we introduced a new framework for low-rank matrix estimation that works by transforming noise models into regularizers via a bootstrap scheme. Our method can adapt to non-isotropic noise structures, thus enabling it to substantially outperform its competitors on problems with, e.g., Poisson noise.

At a high level, our framework works by creating pseudo-datasets $\tX$ from $X$ using the bootstrap distribution $\tlaw\p{X}$. If two pseudo-datasets $\tX_1$ and $\tX_2$ are both likely given $\tlaw\p{X}$, then we want the induced mean estimates $\tilde{\mu}^1_k = \tX_1 \hB_k$ and $\tilde{\mu}^2_k = \tX_2 \hB_k$ to be close to each other. The stable autoencoder \eqref{eq:autoencoder_boot} enables us to turn this intuition into a concrete regularizer by using the L\'evy bootstrap. It remains to be seen whether this idea of regularization via bootstrapping pseudo-datasets can be extended to other classes of low-rank matrix algorithms, e.g., those discussed by \citet{collins2001ageneralization}, \citet{deLeeuw2006PCA}, or \citet{udell2014generalized}.

\section{Appendix: Proofs}

\subsection{Proof of Theorem \ref{theo:gauss}}
We begin by establishing the equivalence between \eqref{eq:autoencoder_boot} and \eqref{eq:ridge}. By bias-variance decomposition, we can check that
\begin{align*}
\EE[\varepsilon_{ij} \simiid \nn\p{0, \, \sigma^2}]{\Norm{X - \p{X + \varepsilon}B}_2^2}
&= \Norm{X - XB}_2^2 + \EE[\varepsilon]{\Norm{\varepsilon B}^2_2} \\
&= \Norm{X - XB}_2^2 + \sum_{i, j, k} \Var{\varepsilon_{ij}} B_{jk}^2 \\
&=  \Norm{X - XB}_2^2 + n \, \frac{\delta\,\sigma^2}{1 - \delta}  \Norm{B}_2^2,
\end{align*}
and so the two objectives are equivalent.

To show that $\hmu_k^\stable$ can be written as \eqref{eq:gauss_thm}, we solve for $\hmu^\stable_k = X\hB_k$ explicitly, where
$$  \hB_k = \argmin_B \left\{\Norm{X - XB}_2^2 + \lambda \Norm{B}_2^2 : \rank{B} \leq k\right\}. $$
Let $ X = UDV^\top$ be the SVD of $X$. 
For any matrix $M$ of the same dimension as $D$, $||UMV^\top||_2^2 = ||M||_2^2$. Thus, we can equivalently write the problem
$$ \hB_k = V\widehat{Q}_k V^\top, \text{ where } \widehat{Q}_k = \argmin_{Q} \left\{\Norm{D -  D Q}_2^2 + \lambda\Norm{Q}_2^2 : \rank{Q} \leq k\right\}. $$
Now, because $D$ is diagonal, $ (DQ )_{ij} = D_{ii} Q_{ij} $.
Thus, we conclude that $\widehat{Q}_{ij} = 0$ for all $i \neq j$, while the problem separates for all the diagonal terms. Without the rank constraint on $Q$, we find that the diagonal terms $\widehat{Q}_{ii}$ are given by
$$\widehat{Q}_{ii} = \argmin_{Q_{ii}} \left\{ \p{1 - Q_{ii}}^2 D_{ii}^2 + \lambda \, Q_{ii}^2 \right\}  = \frac{D_{ii}^2}{\lambda + D_{ii}^2}. $$
Meanwhile, we can check that adding the rank constraint amounts to zeroing out all but the $k$ largest of the $\widehat{Q}_{ii}$.
Thus, plugging this into our expression of $\hmu$, we get that
$$ \hmu_k^\stable= \sum_{i = 1}^k U_{i.} \frac{D_{ii}}{1 + \lambda/D_{ii}^2} V_{i.}^\top. $$

\subsection{Proof of Theorem \ref{theo:compute}}
We start by showing that $\hB$ is the solution to the unconstrained version of \eqref{eq:nongauss}.
Let $V$ be a matrix defined by
$$ V_{ij} = \Var[\tX \sim \tlaw\p{X}]{\tX_{ij}}. $$
Because $\tX$ has mean $X$, we can check that
$$ \EE[\tX \sim \tlaw\p{X}]{\Norm{X - \tX B}_2^2} = \Norm{X - XB}_2^2 + \sum_{i,j,k} V_{ij} \, B^2_{jk}. $$
Thus,
$$\frac12 \, \frac{\partial}{\partial B_{jk}} \EE[\tX \sim \tlaw\p{X}]{\Norm{X - \tX B}_2^2} = - \sum_{i} X_{ij} \p{X - XB}_{ik} + \sum_i V_{ij} \, B_{jk}. $$
Setting gradients to zero, we find an equilibrium
$$ X^\top X =  X^\top X B + S B, \; \where S_{jk} = \begin{cases} \sum_{i = 1}^n V_{ij} &\text{for } j = k, \\ 0 &\text{else.} \end{cases}  $$
Thus, we conclude that
$$ \hB = \p{X^\top X + S}^{-1} \, X^\top X $$
is in fact the solution to \eqref{eq:nongauss} without the rank constraint.

Next, we show how we can get from $\hB$ to $\hB_k$ using \eqref{eq:nongauss_svd}. For any matrix $B$, we can verify by quadratic expansion that
\begin{align*}
\Norm{X - XB}_2^2
&= \Norm{X - X \hB}_2^2 + \Norm{X\p{\hB - B}}_2^2 + 2 \tr\p{\p{X - X\hB}^\top \p{X \hB - X B}} \\
&= \Norm{X - X \hB}_2^2 + \Norm{X\p{\hB - B}}_2^2 \\
&\ \ \ \ \ \ + 2 \tr\p{\p{ X^\top X \p{X^\top X + S}^{-1} X^\top X - X^\top X} \p{B - \hB}}
\end{align*}
Meanwhile,
\begin{align*}
\Norm{S^\frac12 B}^2
&= \Norm{S^\frac12 \hB}_2^2 + \Norm{S^\frac12 \p{B - \hB}}_2^2 + 2 \tr\p{\hB^\top S \p{B - \hB}} \\
&= \Norm{S^\frac12 \hB}_2^2 + \Norm{S^\frac12 \p{B - \hB}}_2^2 + 2 \tr\p{X^\top X \p{X^\top X + S}^{-1} S \p{B - \hB}}.
\end{align*}
Summing everything together, we find that
$$ \Norm{X - XB}_2^2 + \Norm{S^\frac12 B}^2 = \Norm{X\p{B - \hB}}_2^2 + \Norm{S^\frac12 \p{B - \hB}}_2^2 + R\p{\hB, \, X} $$
where $R$ is a residual term that does not depend on $B$. Thus, we conclude that
\begin{align*}
\hB_k
&= \argmin_B \left\{\Norm{X - XB}_2^2 + \Norm{S^{\frac12}B}_2^2 : \rank{B} \leq k\right\} \\
&= \argmin_B \left\{ \tr\p{\p{B - \hB}^\top \p{X^\top X + S} \p{B - \hB}} : \rank{B} \leq k\right\}.
\end{align*}
As shown in, e.g., \citet{takane2013gsvd}, we can solve this last problem by taking the top $k$ terms of the eigendecomposition of $\hB^\top \p{X^\top X + S}^{-1} \hB$.

\subsection{Proof of Theorem \ref{theo:iterate}}
For iterates $t = 0, \, 1, \, ...$, define $M_t = \hmu_t^\top \hmu_t$. Here, we will show that $M_t$ converges to a fixed point $M^*$, and that $M^* \preceq X^\top X$; the desired conclusion then follows immediately.
First, by construction, we have that
$$ M_0 = X^\top X \; \eqand \; M_1 = X^\top X \p{ X^\top X + S}^{-1} X^\top X \p{X^\top X + S}^{-1} X^\top X, $$
and so we immediately see that $M_1 \preceq M_0$.
The general update for $M_t$ is
\begin{equation}
\label{eq:Mk}
M_{t + 1} = g\p{M_t}^\top g\p{M_t}, \; \where g\p{M} = \Sigma^\frac12 \p{M + S}^{-1} M,
\end{equation}
where $\Sigma^\frac12$ is a positive semi-definite solution to $(\Sigma^\frac12)^\top \Sigma^\frac12 = X^\top X$.
Now, because matrix inversion is a monotone decreasing function over the positive semi-definite cone and $S \succ 0$, we find that
$$ g(M) = \Sigma^\frac12 \p{ I - \p{M + S}^{-1} S} $$
is monotone increasing in $M$ over the positive semi-definite cone. In particular
$$ \text{if} \; M_t \preceq M_{t - 1}, \; \text{then} \; M_{t + 1} \preceq M_t. $$
By induction, the sequence $M_t$ is monotone decreasing with respect to the positive semi-definite cone order; by standard arguments, it thus follows that this sequence must converge to a limit $M^*$.
Finally, we note that convergence of $M_t$ also implies convergence of $\hmu_t$, since $\hmu_{t+1} = X\hB_t$ and $\hB_t$ only depends on $\hmu_t$ through $M_t$.

\subsection{Proof of Theorem \ref{theo:low_rank}}
As in the proof of Theorem \ref{theo:iterate}, let $M^* = \hmu^\top \hmu$. Because $M^*$ is a fixed point, we know that
$$ M^* = M^* \p{M^* + S}^{-1} X^\top X \p{M^* + S}^{-1} M^*. $$
Now, because $M^*$ is symmetric with eigenvector $u$, we can decompose it as
$$ M^* = M^\perp + \lambda_u u u^\top, \; \where \; \lambda_u = u^\top M^* u \; \eqand \; \Norm{M^\perp u}_2 = 0.$$
By combining these equalities and using the monotonicity of matrix inversion, we find that
\begin{align*}
\lambda_u
&= u^\top M^* u \\
&= \lambda_u^2 \, u^\top \p{M^* + S}^{-1} X^\top X \p{M^* + S}^{-1} u \\
&\leq \lambda_u^2 \, u^\top S^{-1} X^\top X S^{-1} u.
\end{align*}
This relation can only hold if $\lambda_u = 0$, or $1 \leq \lambda_u \,  u^\top S^{-1} X^\top X S^{-1} u$, and so our desired conclusion must hold.

\subsection{Proof of Proposition \ref{prop:iter_gauss}}
First, we note that in the setting \eqref{eq:gauss} with $\delta = 1/2$, we have $S = n \sigma^2 I$.
Using Theorem \ref{theo:gauss}, we can verify that the singular vectors of $\hmu_t$ are the same as
those of $X$ for each iterate $t = 1, \, 2, \, ... $ of our algorithm; and so we can write the limit
$\hmu^{\iter}$ as singular value shrinker
$$ \hmu^{\iter} = \sum_{l = 1}^{\min\{n, \, p\}} u_l \, \psi\p{d_l} \, v_l^\top. $$
It remains to derive the form of $\psi$. Now, using \eqref{eq:gauss_thm}, we can verify
that the fixed point condition on $\hmu^{\iter}$ can be expressed in terms of $\psi$ as
$$ \psi(d) = \frac{\psi^2(d)}{\lambda + \psi^2(d)} \, d, \ \text{ with } \ \lambda = n\sigma^2. $$
This is a cubic equation, with solutions
$$ \psi(d) = 0, \ \eqand \ \psi(d) = \frac{1}{2}\p{d \pm \sqrt{d^2 - 4n\sigma^2}} \ \eqfor \ d^2 \geq 4 n \sigma^2. $$
Finally, we can verify that our iterative procedure cannot jump over the largest root, thus resulting in
the claimed shrinkage function.

We also check here that, in the case $n = p$, our shrinker $\psi(\cdot)$ is equivalent to the asymptotically
optimal shrinker for operator loss $\psi^*_{op}(d)$ \citep{gavish2014optimalshrink} which is 0
for $d^2 < 4 n \sigma^2$, and else
\begin{align*}
\psi_{op}(d)
&= \frac{1}{\sqrt{2}} \,  \sqrt{d^2 - 2n\sigma^2 + \sqrt{\p{d^2 - 2n\sigma^2}^2 - 4n^2\sigma^4}} \\
&= \frac{1}{\sqrt{2}} \,  \sqrt{d^2 - 2n\sigma^2 + d \, \sqrt{d^2 - 4n\sigma^2}}.
\end{align*}
Squaring our iterative shrinker, we see that
$$ \psi^2(d) = \frac{1}{2}\p{d^2 - 2n\sigma^2 + d\sqrt{d^2 - 4n\sigma^2}} = \psi^2_{op}(d) $$
for $d^2 \geq 4 n \sigma^2$, and 0 else.

\section*{Acknowledgment}

The authors are grateful for helpful feedback from David Donoho, Bradley Efron, William Fithian and Jean-Philippe Vert, as well as two anonymous referees and the JMLR action editor. Part of this work was performed while J.J. was visiting Stanford University, with support from an AgreenSkills fellowship of the European Union Marie-Curie FP7 COFUND People Programme. S.W. was partially supported by a B.C. and E.J. Eaves Stanford Graduate Fellowship.

\spacingset{\SPACESMALL}
\bibliography{references}

\begin{thebibliography}{70}
\providecommand{\natexlab}[1]{#1}
\providecommand{\url}[1]{\texttt{#1}}
\expandafter\ifx\csname urlstyle\endcsname\relax
  \providecommand{\doi}[1]{doi: #1}\else
  \providecommand{\doi}{doi: \begingroup \urlstyle{rm}\Url}\fi

\bibitem[Baldi and Hornik(1989)]{baldi1989neural}
Pierre Baldi and Kurt Hornik.
\newblock Neural networks and principal component analysis: Learning from
  examples without local minima.
\newblock \emph{Neural Networks}, 2\penalty0 (1):\penalty0 53--58, 1989.

\bibitem[Baldi and Sadowski(2014)]{baldi2014dropout}
Pierre Baldi and Peter Sadowski.
\newblock The dropout learning algorithm.
\newblock \emph{Artificial Intelligence}, 210:\penalty0 78--122, 2014.

\bibitem[Benz\'ecri(1973)]{Benz73}
{J-P} Benz\'ecri.
\newblock \emph{L'analyse des donn\'ees. Tome II: L'analyse des
  correspondances}.
\newblock Dunod, 1973.

\bibitem[Benz\'ecri(1986)]{Benz69}
J.-P. Benz\'ecri.
\newblock Statistical analysis as a tool to emerge patterns from the data.
\newblock In S.~Watanabe (ed), editor, \emph{Methodologies of Pattern
  recognition}, pages 34--74. New-York, Academic Press, 1986.

\bibitem[Bishop(1995)]{bishop1995training}
Chris~M Bishop.
\newblock Training with noise is equivalent to {T}ikhonov regularization.
\newblock \emph{Neural Computation}, 7\penalty0 (1):\penalty0 108--116, 1995.

\bibitem[Blei et~al.(2003)Blei, Ng, and Jordan]{blei2003latent}
David~M Blei, Andrew~Y Ng, and Michael~I Jordan.
\newblock Latent {D}irichlet allocation.
\newblock \emph{The Journal of Machine Learning Research}, 3:\penalty0
  993--1022, 2003.

\bibitem[Bourlard and Kamp(1988)]{bourlard1988auto}
Herv{\'e} Bourlard and Yves Kamp.
\newblock Auto-association by multilayer perceptrons and singular value
  decomposition.
\newblock \emph{Biological Cybernetics}, 59\penalty0 (4-5):\penalty0 291--294,
  1988.

\bibitem[Breiman(1996)]{breiman1996bagging}
Leo Breiman.
\newblock Bagging predictors.
\newblock \emph{Machine learning}, 24\penalty0 (2):\penalty0 123--140, 1996.

\bibitem[Buja and Stuetzle(2006)]{buja2006observations}
Andreas Buja and Werner Stuetzle.
\newblock Observations on bagging.
\newblock \emph{Statistica Sinica}, pages 323--351, 2006.

\bibitem[Buntine(2002)]{buntine2002variational}
Wray Buntine.
\newblock Variational extensions to {EM} and multinomial {PCA}.
\newblock In Tapio Elomaa, Heikki Mannila, and Hannu Toivonen, editors,
  \emph{Machine Learning: ECML 2002}, Lecture Notes in Computer Science, pages
  23--34. Springer Berlin Heidelberg, 2002.

\bibitem[Cai et~al.(2010)Cai, Cand{\`e}s, and Shen]{cai2010singular}
Jian-Feng Cai, Emmanuel~J Cand{\`e}s, and Zuowei Shen.
\newblock A singular value thresholding algorithm for matrix completion.
\newblock \emph{SIAM Journal on Optimization}, 20\penalty0 (4):\penalty0
  1956--1982, 2010.

\bibitem[Cand{\`e}s and Tao(2010)]{candes2010power}
Emmanuel~J Cand{\`e}s and Terence Tao.
\newblock The power of convex relaxation: Near-optimal matrix completion.
\newblock \emph{IEEE Transactions on Information Theory}, 56\penalty0
  (5):\penalty0 2053--2080, 2010.

\bibitem[Cand{\`e}s et~al.(2013)Cand{\`e}s, Sing-Long, and
  Trzasko]{candes2013sure}
Emmanuel~J Cand{\`e}s, Carlos~A Sing-Long, and Joshua~D Trzasko.
\newblock Unbiased risk estimates for singular value thresholding and spectral
  estimators.
\newblock \emph{IEEE Transactions on Signal Processing}, 61\penalty0
  (19):\penalty0 4643--4657, 2013.

\bibitem[Chatterjee(2015)]{Chat2014univ}
Sourav Chatterjee.
\newblock Matrix estimation by universal singular value thresholding.
\newblock \emph{The Annals of Statistics}, 43\penalty0 (1):\penalty0 177--214,
  2015.

\bibitem[Collins et~al.(2001)Collins, Dasgupta, and
  Schapire]{collins2001ageneralization}
Michael Collins, Sanjoy Dasgupta, and Robert~E. Schapire.
\newblock A generalization of principal component analysis to the exponential
  family.
\newblock In \emph{Advances in Neural Information Processing Systems}. MIT
  Press, 2001.

\bibitem[d'Aspremont et~al.(2012)d'Aspremont, Bach, and
  El~Ghaoui]{d2012approximation}
Alexandre d'Aspremont, Francis Bach, and Laurent El~Ghaoui.
\newblock Approximation bounds for sparse principal component analysis.
\newblock \emph{Mathematical Programming}, pages 1--22, 2012.

\bibitem[de~Leeuw(2006)]{deLeeuw2006PCA}
Jan de~Leeuw.
\newblock Principal component analysis of binary data by iterated singular
  value decomposition.
\newblock \emph{Computational Statistics and Data Analysis}, 50\penalty0
  (1):\penalty0 21--39, 2006.

\bibitem[Deerwester et~al.(1990)Deerwester, Dumais, Landauer, Furnas, and
  Harshman]{deerwester1990indexing}
Scott~C. Deerwester, Susan~T Dumais, Thomas~K. Landauer, George~W. Furnas, and
  Richard~A. Harshman.
\newblock Indexing by latent semantic analysis.
\newblock \emph{Journal of the American Society for Information Science},
  41\penalty0 (6):\penalty0 391--407, 1990.

\bibitem[Durrett(2010)]{durrett2010probability}
Rick Durrett.
\newblock \emph{Probability: Theory and Examples}.
\newblock Cambridge University Press, 2010.

\bibitem[Efron(2012)]{efron2012large}
Bradley Efron.
\newblock \emph{Large-Scale Inference: Empirical Bayes Methods for Estimation,
  Testing, and Prediction}.
\newblock Cambridge University Press, 2012.

\bibitem[Efron and Tibshirani(1993)]{efron1993introduction}
Bradley Efron and Robert Tibshirani.
\newblock \emph{An Introduction to the Bootstrap}.
\newblock Chapman \& Hall/CRC, 1993.

\bibitem[Escoufier(1973)]{escoufier1973RV}
Yves Escoufier.
\newblock Le traitement des variables vectorielles.
\newblock \emph{Biometrics}, 29:\penalty0 751--760, 1973.

\bibitem[Fithian and Mazumder(2013)]{fithian2013scalable}
William Fithian and Rahul Mazumder.
\newblock Scalable convex methods for flexible low-rank matrix modeling.
\newblock \emph{arXiv preprint arXiv:1308.4211}, 2013.

\bibitem[Gavish and Donoho(2014{\natexlab{a}})]{gavish2014optimal}
Matan Gavish and David~L Donoho.
\newblock The optimal hard threshold for singular values is 4/sqrt (3).
\newblock \emph{IEEE Transactions on Information Theory}, 60\penalty0 (8),
  2014{\natexlab{a}}.

\bibitem[Gavish and Donoho(2014{\natexlab{b}})]{gavish2014optimalshrink}
Matan Gavish and David~L Donoho.
\newblock Optimal shrinkage of singular values.
\newblock \emph{arXiv:1405.7511v2}, 2014{\natexlab{b}}.

\bibitem[Globerson and Roweis(2006)]{globerson2006nightmare}
Amir Globerson and Sam Roweis.
\newblock Nightmare at test time: Robust learning by feature deletion.
\newblock In \emph{Proceedings of the International Conference on Machine
  Learning}, 2006.

\bibitem[Goodfellow et~al.(2013)Goodfellow, Warde-farley, Mirza, Courville, and
  Bengio]{goodfellow2013maxout}
Ian Goodfellow, David Warde-farley, Mehdi Mirza, Aaron Courville, and Yoshua
  Bengio.
\newblock Maxout networks.
\newblock In \emph{Proceedings of the 30th International Conference on Machine
  Learning}, pages 1319--1327, 2013.

\bibitem[Goodman(1985)]{good1985analysis}
L.~A. Goodman.
\newblock The analysis of cross-classified data having ordered and/or unordered
  categories: Association models, correlation models, and asymmetry models for
  contingency tables with or without missing entries.
\newblock \emph{Annals of Statistics}, 13:\penalty0 10--69, 1985.

\bibitem[Greenacre(1984)]{green1984ca}
Michael~J Greenacre.
\newblock \emph{Theory and Applications of Correspondence Analysis}.
\newblock Acadamic Press, 1984.

\bibitem[Greenacre(2007)]{green2007ca}
Michael~J Greenacre.
\newblock \emph{Correspondence Analysis in Practice, Second Edition}.
\newblock Chapman \& Hall, 2007.

\bibitem[Hirschfeld(1935)]{hirschfeld1935connection}
Hermann~O Hirschfeld.
\newblock A connection between correlation and contingency.
\newblock \emph{Mathematical Proceedings of the Cambridge Philosophical
  Society}, 31\penalty0 (4):\penalty0 520--524, 1935.

\bibitem[Hofmann(2001)]{hofmann2001unsupervised}
Thomas Hofmann.
\newblock Unsupervised learning by probabilistic latent semantic analysis.
\newblock \emph{Machine Learning}, 42\penalty0 (1-2):\penalty0 177--196, 2001.

\bibitem[Johnstone(2001)]{john2001asympeignull}
Ian Johnstone.
\newblock On the distribution of the largest eigenvalue in principal components
  analysis.
\newblock \emph{The Annals of Statistics}, 29\penalty0 (2):\penalty0 295--327,
  2001.

\bibitem[Jolliffe(2002)]{jolliffe2002pca}
Ian Jolliffe.
\newblock \emph{Principal Component Analysis}.
\newblock Springer, 2002.

\bibitem[Jolliffe et~al.(2003)Jolliffe, Trendafilov, and
  Uddin]{jolliffe2003modified}
Ian~T Jolliffe, Nickolay~T Trendafilov, and Mudassir Uddin.
\newblock A modified principal component technique based on the lasso.
\newblock \emph{Journal of Computational and Graphical Statistics}, 12\penalty0
  (3):\penalty0 531--547, 2003.

\bibitem[Josse and Holmes(2013)]{josse2013measures}
Julie Josse and Susan Holmes.
\newblock Measures of dependence between random vectors and tests of
  independence. {L}iterature review.
\newblock \emph{arXiv preprint arXiv:1307.7383}, 2013.

\bibitem[Josse and Husson(2011)]{josse2011gcvpca}
Julie Josse and Fran{\c{c}}ois Husson.
\newblock Selecting the number of components in {PCA} using cross-validation
  approximations.
\newblock \emph{Computational Statististics and Data Analysis}, 56\penalty0
  (6):\penalty0 1869--1879, 2011.

\bibitem[Josse and Sardy(2015)]{josse2014adaptive}
Julie Josse and Sylvain Sardy.
\newblock Adaptive shrinkage of singular values.
\newblock \emph{Statistics and Computing}, pages 1--10, 2015.

\bibitem[Josse et~al.(2016)Josse, Sardy, and Wager]{josse2016denoiser}
Julie Josse, Sylvain Sardy, and Stefan Wager.
\newblock \texttt{denoiseR}: A package for low rank matrix estimation.
\newblock \emph{arXiv preprint arXiv:1602.01206}, 2016.

\bibitem[Koren et~al.(2009)Koren, Bell, and Volinsky]{koren2009matrix}
Yehuda Koren, Robert Bell, and Chris Volinsky.
\newblock Matrix factorization techniques for recommender systems.
\newblock \emph{Computer}, 42\penalty0 (8):\penalty0 30--37, 2009.

\bibitem[Krizhevsky et~al.(2012)Krizhevsky, Sutskever, and
  Hinton]{krizhevsky2012imagenet}
Alex Krizhevsky, Ilya Sutskever, and Geoffrey~E Hinton.
\newblock Imagenet classification with deep convolutional neural networks.
\newblock In \emph{Advances in Neural Information Processing Systems}, pages
  1097--1105, 2012.

\bibitem[Kucukelbir and Blei(2015)]{kucukelbir2015population}
Alp Kucukelbir and David~M Blei.
\newblock Population empirical {B}ayes.
\newblock \emph{Uncertainty in Artificial Intelligence}, 2015.

\bibitem[L\^e et~al.(2008)L\^e, Josse, and Husson]{hussonfacto2008}
S\'ebastien L\^e, Julie Josse, and Fran\c{c}ois Husson.
\newblock {FactoMineR}: An {R} package for multivariate analysis.
\newblock \emph{Journal of Statistical Software}, 25\penalty0 (1):\penalty0
  1--18, 2008.

\bibitem[Leek and Storey(2007)]{leek2007capturing}
Jeffrey~T Leek and John~D Storey.
\newblock Capturing heterogeneity in gene expression studies by surrogate
  variable analysis.
\newblock \emph{PLoS genetics}, 3\penalty0 (9):\penalty0 e161, 2007.

\bibitem[Li and Tao(2013)]{li2013simple}
J.~Li and D.~Tao.
\newblock Simple exponential family {PCA}.
\newblock \emph{IEEE Transactions on Neural Networks and Learning Systems},
  24\penalty0 (3):\penalty0 485--497, 2013.

\bibitem[Lustig et~al.(2008)Lustig, Donoho, Santos, and
  Pauly]{lustig2008compressed}
Michael Lustig, David~L Donoho, Juan~M Santos, and John~M Pauly.
\newblock Compressed sensing {MRI}.
\newblock \emph{Signal Processing Magazine, IEEE}, 25\penalty0 (2):\penalty0
  72--82, 2008.

\bibitem[Ng et~al.(2002)Ng, Jordan, and Weiss]{ng2002spectral}
Andrew~Y Ng, Michael~I Jordan, and Yair Weiss.
\newblock On spectral clustering: Analysis and an algorithm.
\newblock \emph{Advances in Neural Information Processing Systems}, 2:\penalty0
  849--856, 2002.

\bibitem[Owen and Eckles(2012)]{owen2012bootstrapping}
Art~B Owen and Dean Eckles.
\newblock Bootstrapping data arrays of arbitrary order.
\newblock \emph{The Annals of Applied Statistics}, 6\penalty0 (3):\penalty0
  895--927, 2012.

\bibitem[Pang and Lee(2004)]{pang2004sentimental}
Bo~Pang and Lillian Lee.
\newblock A sentimental education: Sentiment analysis using subjectivity
  summarization based on minimum cuts.
\newblock In \emph{Proceedings of the Association for Computational
  Linguistics}, page 271. Association for Computational Linguistics, 2004.

\bibitem[Price et~al.(2006)Price, Patterson, Plenge, Weinblatt, Shadick, and
  Reich]{price2006principal}
Alkes~L Price, Nick~J Patterson, Robert~M Plenge, Michael~E Weinblatt, Nancy~A
  Shadick, and David Reich.
\newblock Principal components analysis corrects for stratification in
  genome-wide association studies.
\newblock \emph{Nature Genetics}, 38\penalty0 (8):\penalty0 904--909, 2006.

\bibitem[Robbins(1985)]{robbins1985empirical}
Herbert Robbins.
\newblock \emph{The Empirical Bayes Approach to Statistical Decision Problems}.
\newblock Springer, 1985.

\bibitem[Shabalin and Nobel(2013)]{shabalin2013reconstruction}
Andrey~A Shabalin and Andrew~B Nobel.
\newblock Reconstruction of a low-rank matrix in the presence of {G}aussian
  noise.
\newblock \emph{Journal of Multivariate Analysis}, 118:\penalty0 67--76, 2013.

\bibitem[Shi and Malik(2000)]{shi2000normalized}
Jianbo Shi and Jitendra Malik.
\newblock Normalized cuts and image segmentation.
\newblock \emph{Pattern Analysis and Machine Intelligence, IEEE Transactions
  on}, 22\penalty0 (8):\penalty0 888--905, 2000.

\bibitem[Simard et~al.(2000)Simard, Le~Cun, Denker, and
  Victorri]{simard2000transformation}
Patrice~Y Simard, Yann~A Le~Cun, John~S Denker, and Bernard Victorri.
\newblock Transformation invariance in pattern recognition: Tangent distance
  and propagation.
\newblock \emph{International Journal of Imaging Systems and Technology},
  11\penalty0 (3):\penalty0 181--197, 2000.

\bibitem[Srivastava et~al.(2014)Srivastava, Hinton, Krizhevsky, Sutskever, and
  Salakhutdinov]{srivastava2014dropout}
Nitish Srivastava, Geoffrey Hinton, Alex Krizhevsky, Ilya Sutskever, and Ruslan
  Salakhutdinov.
\newblock Dropout: A simple way to prevent neural networks from overfitting.
\newblock \emph{The Journal of Machine Learning Research}, 15\penalty0
  (1):\penalty0 1929--1958, 2014.

\bibitem[Takane(2013)]{takane2013gsvd}
Yoshio Takane.
\newblock \emph{Constrained Principal Component Analysis and Related
  Techniques}.
\newblock Chapman \& Hall, 2013.

\bibitem[Tibshirani(1996)]{tibshirani1996regression}
Robert Tibshirani.
\newblock Regression shrinkage and selection via the lasso.
\newblock \emph{Journal of the Royal Statistical Society. Series B
  (Methodological)}, pages 267--288, 1996.

\bibitem[Udell et~al.(2014)Udell, Horn, Zadeh, and Boyd]{udell2014generalized}
Madeleine Udell, Corinne Horn, Reza Zadeh, and Stephen Boyd.
\newblock Generalized low rank models.
\newblock \emph{arXiv preprint arXiv:1410.0342}, 2014.

\bibitem[van~der Maaten et~al.(2013)van~der Maaten, Chen, Tyree, and
  Weinberger]{van2013learning}
Laurens van~der Maaten, Minmin Chen, Stephen Tyree, and Kilian~Q Weinberger.
\newblock Learning with marginalized corrupted features.
\newblock In \emph{Proceedings of the International Conference on Machine
  Learning}, 2013.

\bibitem[Verbanck et~al.(2013)Verbanck, Josse, and
  Husson]{verbanck2013regularised}
Marie Verbanck, Julie Josse, and Fran{\c{c}}ois Husson.
\newblock Regularised {PCA} to denoise and visualise data.
\newblock \emph{Statistics and Computing}, pages 1--16, 2013.

\bibitem[Wager et~al.(2013)Wager, Wang, and Liang]{wager2013dropout}
Stefan Wager, Sida Wang, and Percy Liang.
\newblock Dropout training as adaptive regularization.
\newblock In \emph{Advances in Neural Information Processing Systems}, pages
  351--359, 2013.

\bibitem[Wager et~al.(2014)Wager, Fithian, Wang, and Liang]{wager2014altitude}
Stefan Wager, William Fithian, Sida Wang, and Percy Liang.
\newblock Altitude training: Strong bounds for single-layer dropout.
\newblock In \emph{Advances in Neural Information Processing Systems},
  volume~27, pages 100--108, 2014.

\bibitem[Wager et~al.(2016)Wager, Fithian, and Liang]{wager2016data}
Stefan Wager, William Fithian, and Percy Liang.
\newblock Data augmentation via {L}\'evy processes.
\newblock \emph{arXiv preprint arXiv:1603.06340}, 2016.

\bibitem[Wang and Manning(2012)]{wang2012baselines}
Sida Wang and Christopher~D Manning.
\newblock Baselines and bigrams: Simple, good sentiment and topic
  classification.
\newblock In \emph{Proceedings of the Association for Computational
  Linguistics}, pages 90--94. Association for Computational Linguistics, 2012.

\bibitem[Wang et~al.(2013)Wang, Wang, Wager, Liang, and
  Manning]{wang2013feature}
Sida~I Wang, Mengqiu Wang, Stefan Wager, Percy Liang, and Christopher~D
  Manning.
\newblock Feature noising for log-linear structured prediction.
\newblock In \emph{Empirical Methods in Natural Language Processing}, 2013.

\bibitem[Witten et~al.(2009)Witten, Tibshirani, and
  Hastie]{witten2009penalized}
Daniela~M Witten, Robert Tibshirani, and Trevor Hastie.
\newblock A penalized matrix decomposition, with applications to sparse
  principal components and canonical correlation analysis.
\newblock \emph{Biostatistics}, 2009.

\bibitem[Xu et~al.(2003)Xu, Liu, and Gong]{xu2003document}
Wei Xu, Xin Liu, and Yihong Gong.
\newblock Document clustering based on non-negative matrix factorization.
\newblock In \emph{ACM SIGIR Conference on Research and Development in
  Information Retrieval}, pages 267--273. ACM, 2003.

\bibitem[Zhang and Huang(2008)]{zhang2008lasso}
C.~H. Zhang and J.~Huang.
\newblock The sparsity and bias of the lasso selection in high-dimensional
  linear regression.
\newblock \emph{The Annals of Statistics}, 36\penalty0 (4):\penalty0
  1567--1594, 2008.

\bibitem[Zou(2006)]{zou2006adapt}
H.~Zou.
\newblock The adaptive {LASSO} and its oracle properties.
\newblock \emph{Journal of the American Statistical Association}, 101:\penalty0
  1418--1429, 2006.

\bibitem[Zou et~al.(2006)Zou, Hastie, and Tibshirani]{zou2006sparsepca}
H.~Zou, T.~Hastie, and R.~Tibshirani.
\newblock Sparse principal component analysis.
\newblock \emph{Journal of Computational and Graphical Statistics}, 15\penalty0
  (2):\penalty0 265--286, 2006.

\end{thebibliography}
\spacingset{\SPACEBIG}

\end{document}